\documentclass{JFM-FLM_Au}

\sethlcolor{yellow}

\usepackage{color}

\lefttitle{Dipendra Gupta,Vedant Kumar, Johan Larsson and Gregory P. Bewley}
\righttitle{Journal of Fluid Mechanics}

\title{Effects of mean flow skew on turbulent shear layers. Part II. Experimental investigation}

\author{Dipendra Gupta\aff{1}, Vedant Kumar\aff{2}, Johan Larsson\aff{2} \and Gregory P. Bewley\aff{1}}

\affiliation{\aff{1}Sibley School of Mechanical and Aerospace Engineering, Cornell University, Ithaca, NY 14850, USA
\aff{2}Department of Mechanical Engineering, University of Maryland, College Park, MD 20742, USA}

\corresau{Dipendra Gupta, \email{dg535@cornell.edu}}

\begin{document}
\maketitle

\begin{abstract}

Planar turbulent mixing layers, formed by the interactions of two parallel streams with different velocities, have been studied far more than three dimensional (3D) turbulent mixing layers, in which the incoming streams are skewed, and not parallel. Yet many practical shear flows are 3D. Here, we develop and validate an experimental methodology to generate and characterize skewed turbulent mixing layers and to quantify how mean-flow skew modifies mixing layer dynamics. We introduce skew with a spanwise deflection of the mean flow using turning vanes mounted near the trailing edge of a splitter plate, and we use cross-wire anemometry to investigate the downstream evolution of the flow. Relative to the planar configuration, the skewed mixing layer exhibits systematic reductions in both mean and turbulent quantities, with deviations reaching approximately 40\%. Despite these quantitative differences, the fundamental characteristics of the mixing layer remain largely unchanged. Mean-velocity profiles collapse under similarity scaling, shear-layer thicknesses retain approximately linear downstream growth, and Reynolds-stress profiles preserve their characteristic near-Gaussian form. Townsend's structure parameter, which quantifies the efficiency of turbulent momentum transport, remains approximately invariant between the planar and skewed configurations, in contrast to skewed turbulent boundary layers, wherein comparable mean flow skewing reduces the parameter by approximately 30\%. These results indicate that mean flow skew modifies turbulent mixing layers quantitatively while exerting only a secondary influence on their underlying dynamics. This study establishes a controlled experimental framework and empirical benchmark for future investigations of three-dimensional free-shear turbulence.

\end{abstract}



\section{Introduction}
\label{sec:Introduction}

Planar turbulent mixing layers are fundamental free-shear flows and have long served as canonical to investigate the development of shear turbulence \citep{townsend1976structure}. 
In their classical formulation, two nominally parallel streams with initially different but constant mean velocities interact across the interface between the streams \citep{brown1974density,jimenez2025chaos}. 
Owing to this relative geometric simplicity, decades of theoretical, numerical, and experimental studies have established a coherent understanding of planar mixing layers, making them central to the modern framework of shear-flow turbulence \citep{Pope_2000}. 
However, in many realistic flows like geophysical flows, delta wing flows, propulsion systems, turbomachinery, and wakes, shear layers arise under conditions that are essentially three-dimensional (3D), in the sense that the mean velocity vectors of the adjoining streams differ not only in magnitude but also in orientation \citep{fiedler1998three}. 
The interactions of such non-parallel freestreams generates a nonzero mean \textit{spanwise} velocity at interface. 
This additional geometric complexity can fundamentally modify the organization of turbulence, interscale dynamics, entrainment processes, and the evolution of coherent motions. 
Despite its practical relevance, the structure and dynamics of such 3D mixing layers remain comparatively unexplored experimentally, owing in large part to the difficulty of generating controlled flow configurations while simultaneously isolating intrinsic mean-flow 3D effects from facility-dependent influences. 
The present work seeks to address this gap by developing and validating a simple experimental methodology for the systematic generation and investigation of 3D turbulent mixing layers, with particular emphasis on identifying departures from canonical planar behavior. 
Beyond the specific results reported here, the study provides a detailed experimental characterization of these flows and establishes a reliable empirical benchmark for future theoretical, numerical, and experimental investigations of non-planar free-shear turbulence.

Three-dimensionality in shear flows may arise either through externally imposed 3D mean-flow conditions or through intrinsic instabilities of nominally two-dimensional (2D) vortical structures. The former occurs in a wide range of canonical and applied configurations, including jets in crossflow \citep{mahesh2013interaction}, flows over swept or finite wings \citep{riley1998development}, turbulent boundary layers developing in curved passages \citep{peacock2024dissipation}, and mixing layers formed between non-parallel walls \citep{paschereit1989mixing}. 
In such flows, the mean velocity field possesses an inherently 3D nature, introducing directional shear due to mean velocity component in the transverse direction. 
Intrinsic three-dimensionality, by contrast, emerges from secondary instabilities of initially quasi-two-dimensional coherent motions. 
In planar mixing layers, for example, the development of streamwise-oriented “rib” vortices during vortex pairing generates persistent streamwise vorticity and promotes 3D turbulent dynamics \citep{bell1992measurements}. 
The focus of the present work is exclusively on mean-flow three-dimensionality.

As reviewed comprehensively in \citet{johnston1996advances}, the mechanisms responsible for generating three-dimensionality in wall-bounded shear flows may broadly be classified as either shear-driven (viscous-induced) or pressure-driven (inviscid-induced). 
Examples of the former include moving-wall configurations \citep{kiesow2003near}, jets in crossflow \citep{mahesh2013interaction}, and rotating-disk boundary layers \citep{littell1994turbulence}, whereas the latter commonly arise in curved geometries \citep{peacock2024dissipation} and flows over wings \citep{riley1998development}. 
Nevertheless, this distinction is not always physically strict \citep{lozano2020non}. 
Indeed, in 3D turbulent boundary layers, many of the resulting characteristics of turbulence appear to be relatively insensitive to the specific mechanism by which three-dimensionality is introduced \citep{johnston1996advances}. 
In the present study, we impose mean flow three-dimensionality at the interface between the high- and low-speed streams through controlled pressure-driven modifications to the boundary conditions near the splitter-plate trailing edge. 

The influence of mean-flow three-dimensionality has been investigated extensively in turbulent boundary layers (TBLs) \citep{johnston1996advances}. 
Relative to canonical two-dimensional boundary layers, three-dimensional TBLs exhibit several distinctive and, in some respects, counter-intuitive features, including a reduction in the primary Reynolds shear stress relative to the turbulent kinetic energy (often characterized through Townsend’s structure parameter), reduced drag, and a pronounced misalignment between the Reynolds stress tensor and the local mean strain or shear directions \citep{bradshaw1985measurements,johnston1996advances,lozano2020non}. 
These effects are generally attributed to the distortion and reorientation of the near-wall vortical structures responsible for turbulence production, together with a reduction in their streamwise coherence lengths \citep{eaton1995effects,kiesow2003near}. 
More recently, the direct numerical simulations in \citet{lozano2020non} demonstrated that an incompressible turbulent channel flow subjected to an imposed spanwise pressure gradient develops a persistent misalignment between the streamwise-oriented vortex cores and the near-wall flow beneath them. 
This reduces the pressure--strain correlation and weakens the production of Reynolds stresses. 
These observations motivate the present investigation: namely, whether analogous modifications to turbulence statistics also arise in turbulent free-shear layers subjected to an externally imposed three-dimensional mean-flow. 

Most investigations of turbulent free-shear layers are focused on canonical 2D mixing layers \citep{bell1990development,brown1974density,druault2005experimental,fiscaletti2016scale,liepmann1947investigations,loucks2012velocity,mcmullan2015organised,rogers1994direct}. 
These studies have established the current understanding of the asymptotic evolution of the mean flow, turbulence statistics, and coherent structure dynamics in shear layers. 
In particular, they have demonstrated that planar mixing layers evolve asymptotically toward a self-preserving state characterized by: (i) linear growth of the mixing-layer thickness with downstream distance, (ii) invariance of the normalized mean velocity profile, and (iii) collapse of the normalized Reynolds-stress profiles together with saturation of their peak values with downstream distance~\citep{townsend1976structure}. 
This asymptotic behavior has provided a framework to model and interpret free-shear turbulence. 

Despite this progress, the transient development preceding the self-preserving state remains comparatively poorly understood, even in nominally planar mixing layers. 
In practice, 3D effects are often already present during these early stages and may themselves evolve transiently before the flow approaches a statistically self-preserving regime. 
While previous studies have primarily examined how different inflow conditions influence the downstream aymptotic state~\citep{bell1990development,d2013organized,mcmullan2015organised,sharan2019turbulent}, 
considerably less attention has been devoted to the transient dynamics themselves, likely because such behavior is strongly configuration-dependent and appears to lack universality. 
Yet, from both engineering and modeling perspectives, understanding the near-field evolution of the flow is essential, since this region governs the initial development of turbulence, entrainment, momentum transport, and ultimately the resulting drag and mixing characteristics. 
The prediction of the transient dynamics is even more challenging when three-dimensionality of the mean flow is introduced, possibly due to the additional coupling between directional shear, coherent structures, and Reynolds-stress evolution. 

Unlike turbulent boundary layers, whose near-wall dynamics are dominated by streamwise-oriented vortical motions \citep{smits2011high}, turbulent mixing layers are characterized during their early and transitional stages by large spanwise-oriented vortex rollers \citep{brown1974density}. 
Since the seminal observations of these coherent structures in planar mixing layers in \citet{brown1974density}, extensive efforts have sought to understand their formation, evolution, and breakdown \citep{d2013organized,rogers1994direct,roshko2005plane,sharan2019turbulent}. 
These structures are argued to arise from the Kelvin--Helmholtz instability developing at the interface between the high- and low-speed streams. 
As the flow evolves downstream, however, the initially quasi-2D rollers become susceptible to a range of secondary mechanisms that generate 3D motion, including spanwise instabilities \citep{chandrsuda1978effect}, intrinsic secondary instabilities \citep{jimenez1985perspective}, externally imposed perturbations \citep{nygaard1991evolution}, or combinations thereof. 
The spanwise instability could deform the primary spanwise rollers, inducing spiral pairing before their complete breakdown while the secondary instability generate streamwise-oriented vortices embedded within the braid regions connecting adjacent rollers. 
These streamwise vortices may subsequently evolve into the helical rib structures further downstream \citep{jimenez1985perspective}. 
The three-dimensionality generated by these intrinsic instabilities is fundamentally distinct from the externally imposed three-dimensionality of the mean flow considered in the present work \citep{fiedler1998three}. 
For detailed discussions of instability-induced three-dimensionality in mixing layers, the reader is referred to the review in \citet{ho1984perturbed}. 
Here, we focus exclusively on mixing layers subjected to externally imposed mean-flow three-dimensionality, which we hereinafter refer to as \textit{skewed mixing layers}.


In contrast to planar mixing layers, comparatively few studies have examined skewed mixing layers \citep{fiedler1998three}. 
One of the earliest experimental investigations is reported in \citet{hackett1970three}, which describes the downstream development of two equal-speed air streams intersecting at $\pm 45^\circ$ relative to a common centerline downstream of a splitter plate. 
Using measurements made with a five-hole probe, a mean-flow analysis reveals that the maximum Reynolds shear stress increased by approximately $40\%$ in the skewed case relative to the planar case. 
Subsequently, \citet{fric1996skewed} investigates skewed mixing layers in a water tunnel for varying velocity ratios and skew angles. 
Measurements of passive-scalar concentration demonstrated that imposed skew enhanced the degree of mixedness, particularly when the more strongly skewed stream also corresponded to the higher-speed stream. 
Similarly, \citet{grundel1993mixing} examines skewed mixing layers with equal freestream velocities and skew angles of $\pm 15^\circ$, reporting that the mean spanwise velocity profile exhibited an approximately hyperbolic-tangent form and that the mixing-layer thickness increased linearly with downstream distance. 
Furthermore, the dominant coherent structures assumed helical forms with vortex axes inclined relative to the splitter-plate trailing edge. 
Comparable conclusions are reported in \citet{schroder2000trailing}, which suggests that the growth rates of skewed mixing layers with velocity ratios between $0.5$ and $1$ and skew angles up to $\pm 20^\circ$ are similar to those of planar mixing layers, while the turbulent kinetic energy increased under skewed conditions, which, they argued, would enhance the mixing. 
The studies described above primarily considered configurations in which the imposed mean flow three-dimensionality generated an additional mean velocity component in the spanwise direction. 
In contrast, \citet{azim2003plane} investigates skewed mixing layers in which the imposed skew acted within the vertical plane, thereby introducing an additional mean vertical velocity component rather than a spanwise one. 
For velocity ratios ranging from $0.7$ to $9$ and skew angles up to $\pm 9^\circ$, they observed that the skewed mixing layers retained several canonical features of planar mixing layers, including linear growth and asymptotic self-similarity. 
However, unlike the conclusions proposed in \citet{hackett1970three}, \citet{azim2003plane} reports a substantial reduction in the normalized peak Reynolds shear stress, $\overline{u'v'}_{\max}/\Delta U^2$, by approximately $75$--$90\%$ relative to the planar case. 
These contrasting observations highlight the limited and, at times, inconsistent understanding currently available regarding the effects of imposed mean-flow three-dimensionality on turbulent mixing layers.

Despite the prevalence of skewed shear layers in both natural and engineering flows, experimental investigations remain scarce, owing in part to the substantial challenges associated with generating controlled three-dimensional free-shear configurations. 
All above studies have relied on facilities employing multiple contraction sections to produce the skewed streams, with the resulting mixing layers typically evolving in an unconfined ambient environment. 
An exception is the work in \citet{azim2003plane}, which examines the flow within the test section of an open-circuit wind tunnel. 
Such complex experimental configurations often introduce strong sensitivity to boundary conditions, increased measurement uncertainty, and difficulties in separating intrinsic flow physics from apparatus-dependent effects. 
These limitations complicate data interpretation, hinder reproducibility, and ultimately restrict the development of a systematic physical understanding of skewed mixing-layer dynamics.

Complementary insight has emerged from theoretical and numerical studies. 
Inviscid stability analyses in \citet{lu1993inviscid} and \citet{lu1999asymptotic} demonstrate that mean-flow three-dimensionality can amplify disturbance growth rates and, at sufficiently large skew angles approaching $90^\circ$, promote vortex breakdown. 
More recently, direct numerical simulations (DNS) in \citet{meldi2020numerical} show that in incompressible skewed mixing layers the influence of imposed skew is most pronounced in the near field, whereas the dynamics further downstream are governed predominantly by the shear intensity. 
These results further indicate that imposed skew enhances the amplification of inlet disturbances and accelerates the development of the mixing region. 
Subsequently, \citet{boukharfane2021direct} presents DNS of compressible skewed mixing layers at a convective Mach number of $0.48$ and reports that imposed skew increases the vorticity thickness, its growth rate, and the magnitudes of the transverse normal, spanwise normal, and primary shear stresses. 
\citet{boukharfane2021direct} attributes these effects to reduced pressure--strain redistribution and enhanced production of the spanwise Reynolds stress component. 
The simulations also reveal strong amplification of inlet disturbances and significant sensitivity of the near-field development to imposed skew. 
It is important to note, however, that both DNS investigations initialized the flow using prescribed hyperbolic-tangent mean velocity profiles with superimposed perturbations to trigger transition. 
Since the form and amplitude of imposed perturbations are known to influence the early evolution of mixing layers \citep{ho1984perturbed}, it remains unclear to what extent the reported transition dynamics arise intrinsically from imposed mean-flow skew rather than from the prescribed velocity profiles and the nature of the superimposed perturbations themselves.

Motivated by the limited experimental literature on skewed mixing layers, the complexity of existing generation techniques, and the lack of consensus among previous observations, the present work develops and demonstrates a simple methodology for generating skewed turbulent mixing layers within a wind tunnel employing a single contraction section. 
Our primary objective is to isolate and investigate the effects of externally imposed mean-flow three-dimensionality on the transient evolution of turbulent mixing layers and to identify departures from canonical planar behavior. 
To this end, we first generate turbulent freestreams with different mean velocities using a classical passive grid partially covered with a mesh, subsequently separate the streams using a splitter plate, and finally impose controlled skew using turning vanes positioned near the splitter-plate trailing edge. 
We then characterize the downstream evolution of the flow using X-wire probes, enabling detailed examination of the transient development of the mean flow and turbulent stresses. 
Beyond the specific findings reported here, the present study provides a systematic experimental characterization of skewed turbulent mixing layers and establishes a reliable empirical benchmark for future theoretical, numerical, and experimental investigations of three-dimensional free-shear turbulence.

As part of a collaborative effort supported by the Office of Naval Research, a companion large eddy simulation (LES) study was performed in Prof. Larsson’s group at the University of Maryland, College Park, to investigate the effects of imposed mean-flow three-dimensionality on turbulent mixing layers. Details regarding the numerical formulation, initialization procedure, and computational setup are provided in the companion LES paper.

The remainder of the paper is organized as follows. 
\S~\ref{sec:Methodology} describes the experimental methodology and facility, including including the identification of a downstream region in which the flow evolution is minimally influenced by apparatus-dependent effects, and a brief overview of the companion LES setup. 
\S~\ref{sec:MeanFlowQuantities} and \S~\ref{sec:turbulentquantities} present the experimental results and discussion, focusing on the downstream relaxation of the mean strain and the development of the turbulent stresses. 
Finally, the main conclusions of the study are summarized in \S~\ref{sec:conclusion}.

\section{Methodology}\label{sec:Methodology}
\subsection{Experiment}\label{sec:Experiment}

\subsubsection{Experimental setup}
\label{sec:ExptSetup}

\begin{figure}

 \centering
 \includegraphics[width=1\linewidth, trim={20 75 15 50},clip]{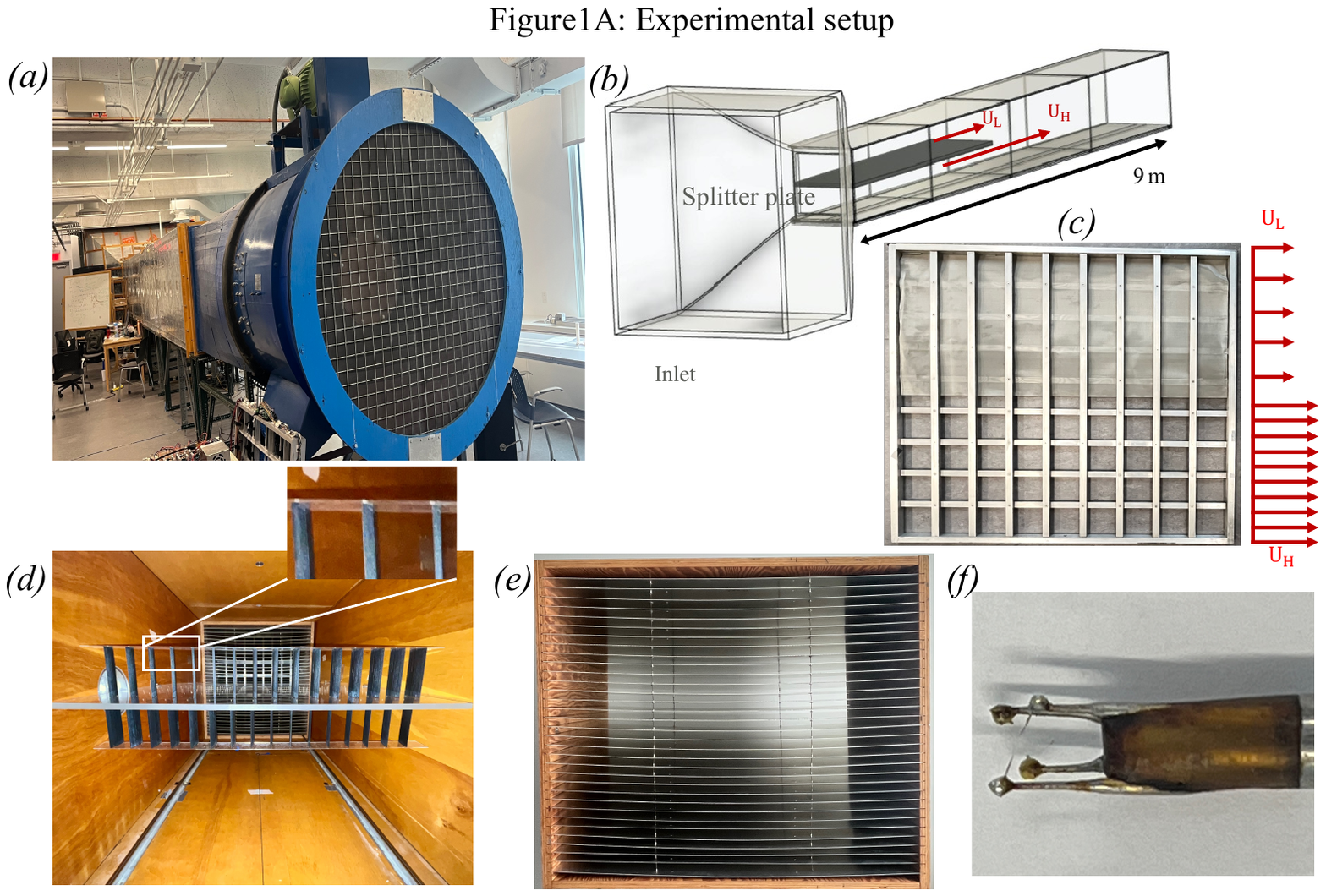}
\caption{Experimental setup for skewed mixing layers. \textit{(a)} Warhaft wind tunnel. \textit{(b)} Isometric view showing splitter plate with high- and low-speed streams at the trailing edge, denoted by $U_H$ and $U_L$, respectively.   \textit{(c)} Mesh grid used to generate different-speed mixing layers. \textit{(d)} Front view of turning-vane and end plates arrangement on the splitter plate. Zoomed-in view shows the end plates mounted on outer edges of the vanes on the low-speed side. \textit{(e)} Flow straightener. \textit{(f)} X-wire thermal anemometer for velocity measurements.}
 \label{fig:Figure1}
\end{figure}

\begin{figure}
 \centering
 \includegraphics[width=1\linewidth, trim={20 315 15 85},clip]{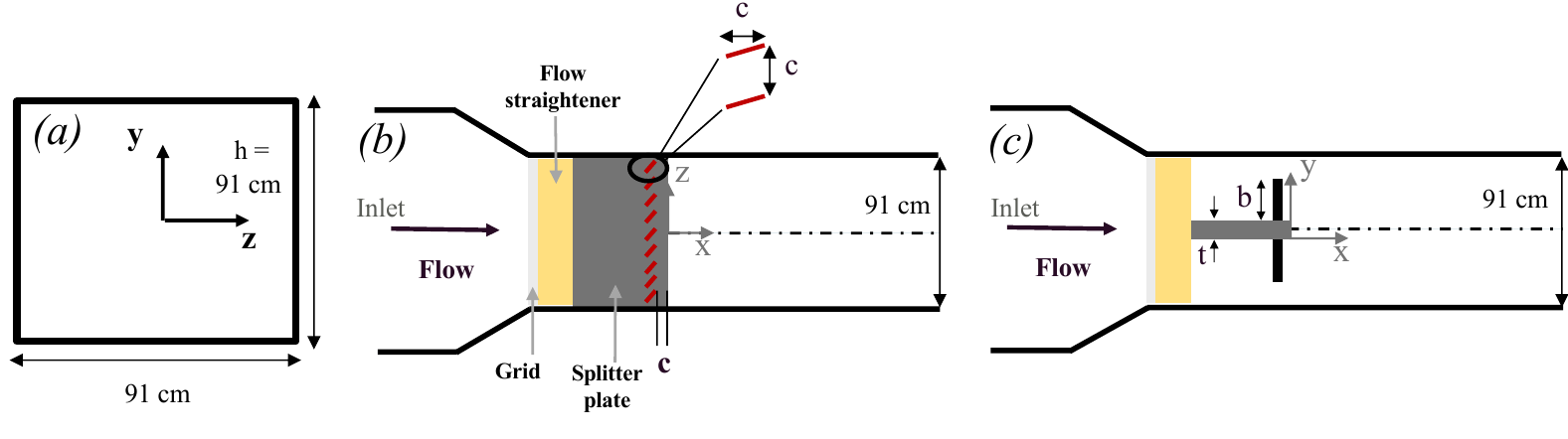}
\caption{Coordinate system.  Schematic (not to scale) showing \textit{(a)} Cross-section. \textit{(b)} Top view showing turning vanes located $\sim 1c$ upstream of the splitter-plate trailing edge, where $c$ is the vane chord length. Sixteen vanes are mounted on each side of the splitter plate; only nine are shown for visual clarity. \textit{(c)} Side view showing turning vanes height, $b$ and splitter plate thickness, $t$.}
 \label{fig:Figure2}
\end{figure}

We conducted the experiments in the Warhaft Wind and Turbulence Tunnel (WWTT) at Cornell University~\citep{yoon1990evolution}. 
The WWTT has a test section with a cross-section of $0.91~\mathrm{m} \times 0.91~\mathrm{m}$ and length $9.1~\mathrm{m}$, 
and freestream velocities up to approximately $20~\mathrm{m/s}$ (figure~\ref{fig:Figure1}(\textit{a, b}) and figure~\ref{fig:Figure2}). 
We generated turbulence using a classical passive square-mesh grid of mesh size $0.083~\mathrm{m}$ placed at the entrance of the test section. 
To produce two parallel turbulent streams with different mean velocities, we covered the upper half of the grid with a stainless-steel wire cloth. 
The additional blockage reduced the velocity in the upper half of the tunnel and generated a low-speed stream above a high-speed stream (figures~\ref{fig:Figure1}(\textit{c}) and \ref{fig:Figure2}(\textit{b,c})). 
We separated the two streams using an acrylic splitter plate of thickness $1.27 \times 10^{-2}~\mathrm{m}$ and with dimensions approximately $0.9~\mathrm{m} \times 1.2~\mathrm{m}$ mounted along the tunnel’s horizontal symmetry plane using supports attached to the side walls (figures~\ref{fig:Figure1}(\textit{b}) and \ref{fig:Figure2}(\textit{b,c})). 
Because of a tendency the flow to deflect from the low-speed side toward the high-speed side, we installed a 0.6\,m long flow straightener (composed of horizontal sheet metal plates spaced by 0.025\,m) between the turbulence-generating grid and the splitter plate (figures~\ref{fig:Figure1}(\textit{e}) and \ref{fig:Figure2}(\textit{b,c})).

To generate skewed mixing layers, we mounted turning vanes with a NACA 0012 profile on both sides of the splitter plate near its trailing edge (figure~\ref{fig:Figure1}(\textit{d})). 
Each vane had chord length $c = 5~\mathrm{cm}$ and height $h = 12~\mathrm{cm}$ (figure~\ref{fig:Figure2}(\textit{b,c})). 
We selected the vane height such that it exceeded the boundary-layer thickness developing along either side of the splitter plate, thereby ensuring that the imposed turning affected the full local flow thickness. 
We fixed the vanes to the splitter plate using threaded rods that allowed us to adjust the vane angles independently with an uncertainty of $\pm 0.2^\circ$. 
We attached acrylic end plates of thickness $0.5\times 10^{-3}~\mathrm{m}$ on the outer edges of the vanes (figure~\ref{fig:Figure1}(\textit{d})). 
These end plates extended downstream by approximately one chord length and spanned the width of the test-section. 
We used 32 vanes in total, with 16 mounted on each side of the splitter plate. 
We spaced the vanes by one chord length and positioned their trailing edges approximately one chord length upstream of the splitter-plate’s trailing edge. 
When we aligned the vanes at $0^\circ$, the configuration reduced to the canonical planar mixing-layer case. 

We measured the velocity field using a rotatable X-wire thermal anemometer mounted on a three-axis traverse system driven by stepper motors. 
The traverse provided a positioning resolution of $2.54 \times 10^{-5}~\mathrm{m}$ in both the cross-stream and spanwise directions. 
Although the streamwise traverse was motor-controlled, it lacked a digital encoder, requiring us to determine downstream positions manually. 
This procedure introduced an estimated positional uncertainty of $\pm 1~\mathrm{cm}$ in the streamwise direction. 
We defined the coordinate system such that $(x,y,z)=(0,0,0)$ corresponded to the splitter-plate trailing edge in the longitudinal symmetry plane on the high-speed side. 
Here, $x$, $y$, and $z$ denote the streamwise, cross-stream, and spanwise directions, respectively. We denote the instantaneous velocity vector by $\boldsymbol{u} (t)=(u_x,u_y,u_z)$ and define the time-averaged velocity as $\textbf{U}= (U, V, W)= 1/T\int \textbf{u} (t)dt$,
where $T$ denotes the total acquisition time at a given spatial location. We define the fluctuating velocity components as $\boldsymbol{u}'(t)=\boldsymbol{u}(t)-\boldsymbol{U}$.

We fabricated the X-wire probes in-house and operated them using a DANTEC StreamLine Pro constant-temperature anemometry (CTA) system. 
Each probe consisted of platinum-coated tungsten wires of diameter $5~\mu\mathrm{m}$, active sensing length approximately $1~\mathrm{mm}$, and wire separation approximately $1~\mathrm{mm}$ (figure~\ref{fig:Figure1}(\textit{f})). 
We calibrated the probes statically in the potential core of a jet and verified selected measurements using a Pitot-static tube. Before digitization, we applied gain and DC offset conditioning to the analog CTA signals and then acquired them using a 16-bit NI USB-6221 data acquisition system. We sampled the signals at $50~\mathrm{kHz}$ for durations corresponding to approximately $10^4$ integral time scales at each measurement location. During post-processing, we corrected the raw voltage signals for the applied gain and offset, and filtered them using a fourth-order low-pass Butterworth filter with a cutoff frequency of $10~\mathrm{kHz}$ to suppress high-frequency noise and minimize aliasing effects.

We acquired measurements in the $xy$-plane at 15 downstream locations spanning $0.02 \le x/h\le 6$, and in $xz$-plane at selected downstream locations. 
At each downstream location, we measured the flow at 18 cross-stream positions within the range $-0.3 \le y/h < 0.3$ in the longitudinal symmetry plane, unless stated otherwise. To assess statistical convergence and uncertainty, we used two independent approaches. First, we applied bootstrap resampling to the full dataset. Second, we subdivided the acquired records into segments corresponding to approximately $10^3$ integral time scales and computed statistics independently for each segment. We estimated the random uncertainty using the combined variance obtained from these two procedures. Based on this analysis, we estimated the uncertainties in the mean velocities, velocity gradients, normal Reynolds stresses, and Reynolds shear stresses to be within approximately $5\%$, $8\%$, $15\%$, and $20\%$, respectively.

For the present study, we fixed the freestream velocities and varied only the turning-vane angles to control the imposed mean flow three-dimensionality. 
Specifically, we set the high- and low-speed freestream velocities to $U_H = 10.5~\mathrm{m/s}$ and $U_L = 4~\mathrm{m/s}$, respectively, corresponding to a velocity ratio $r=U_L/U_H = 0.38$, and a velocity difference $\Delta U = U_H-U_L = 6.5~\mathrm{m/s}$. 
We investigated two vane configurations: (i) $\theta_H=-\theta_L=0^\circ$, corresponding to the planar mixing-layer case, and (ii) $\theta_H=-\theta_L=10^\circ$, corresponding to a skewed mixing layer with an effective relative skew angle of $20^\circ$ between the streams. We define positive turning angles in the counterclockwise sense when viewed in planform looking downstream. 
We selected the velocity ratio to approximately match one of the canonical cases investigated in \citet{brown1974density}, while we limited the turning angle to values below the onset of flow separation over the vanes at the operating Reynolds numbers considered here.

Under these operating conditions, we measured peak streamwise turbulence intensities of approximately $10\%$ in the incoming freestreams, while the transverse turbulence intensities were approximately $5\%$. 
The Reynolds number based on the velocity difference $\Delta U$ and the combined initial momentum thickness was approximately $6500$. 
In addition, the Reynolds number based on the vane chord length and velocity difference was approximately $2\times10^4$, which ensured attached flow over the turning vanes for the imposed deflection angles~\citep{abbott2012theory}. 
Because of experimental constraints, we characterized the inflow conditions approximately $5~\mathrm{mm}$ upstream of the splitter-plate trailing edge rather than directly at the trailing edge itself.

\subsubsection{Statistical convergence}
\label{sec:StatConv}

Turbulent mixing layers contain a broad range of dynamically relevant scales, ranging from the largest energy-containing eddies to the smallest dissipative motions~\citep{Pope_2000}. In shear-dominated flows, the dynamics are further influenced by large-scale coherent structures, such as the spanwise vortex rollers characteristic of mixing layers~\citep{brown1974density}. Reliable estimation of mean and higher-order turbulence statistics therefore requires sufficiently long acquisition times to ensure adequate sampling of both the large-scale coherent motions and the stochastic turbulent fluctuations. 
To assess statistical convergence of the acquired data, we employed three complementary procedures.

\begin{figure}
 \centering
 \includegraphics[width=0.9\linewidth, trim={10 20 25 55},clip]{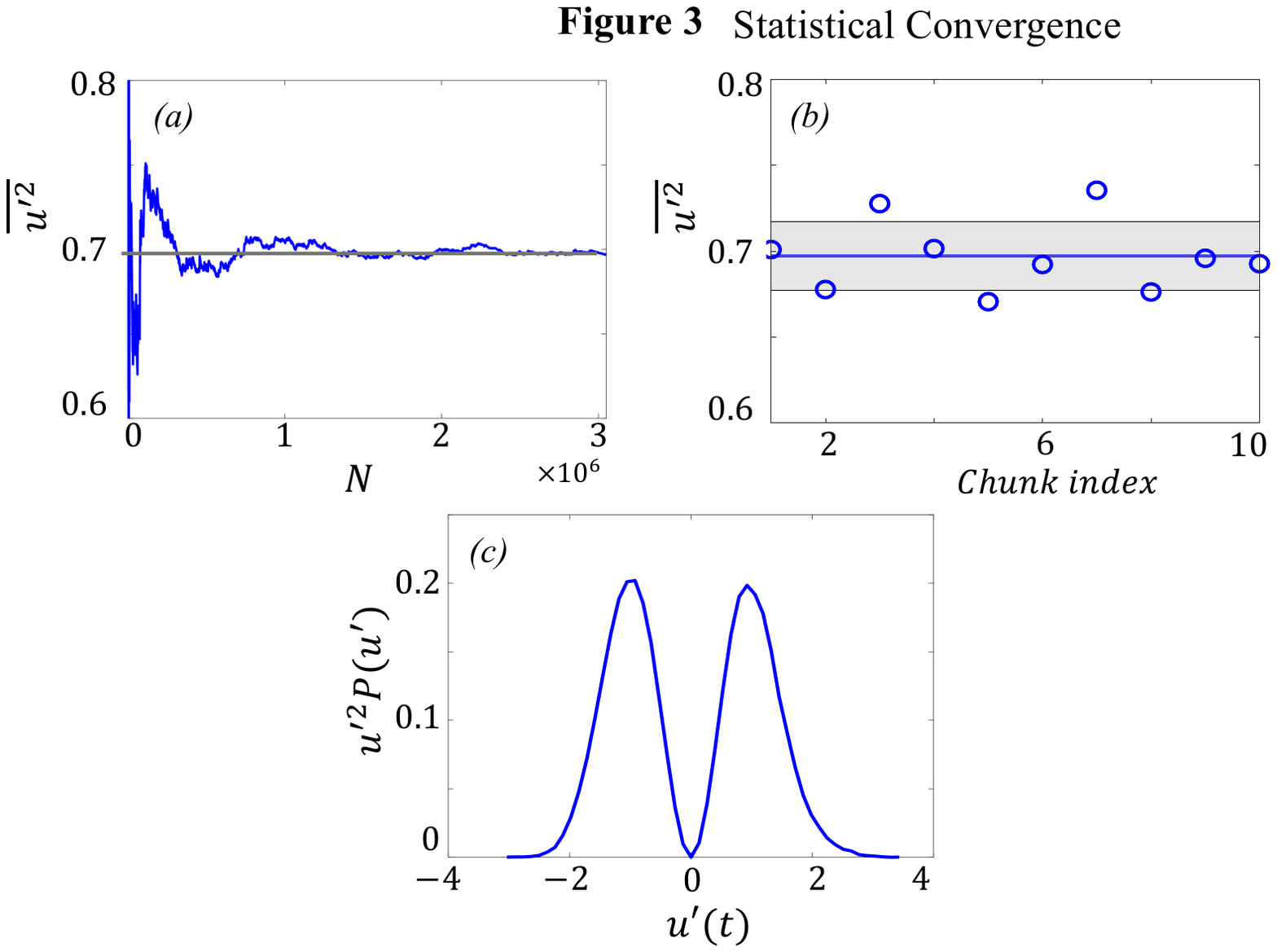}
\caption{Statistical convergence of $\overline{u'^2}$. 
\textit{(a)} Running mean of $\overline{u'^2}$ as a function of the number of samples, $N$. 
\textit{(b)} Chunk-wise time-averaged $u'^2$ (open circles), compared with the value computed from the full acquired signal (blue line). 
The grey shaded region indicates the 95\% confidence interval about the mean of the full signal. 
\textit{(c)} Integrand of the second moment of the probability density function of $u'(t)$. $u'(t)$ is expressed in m/s. }
 \label{fig:Figure3}
\end{figure}

First, we evaluated the running averages of representative turbulent quantities over the entire acquisition record. 
Figure~\ref{fig:Figure3}(\textit{a}) shows the running average of the longitudinal velocity variance, $\overline{u'^2}=\overline{(u(t)-U)^2}$, where the overbar indicates time averaging. At small sample sizes, the running estimate exhibits substantial fluctuations associated with insufficient sampling of the large-scale motions. As the number of samples increases, however, these fluctuations progressively diminish and the estimate approaches an approximately constant asymptotic value. In the present measurements, the running average stabilized for sample sizes exceeding approximately $N\sim10^6$, indicating that the acquisition duration of approximately over $20~\mathrm{s}$ at sampling frequency of $50~\mathrm{kHz}$ was sufficient to obtain statistically converged second-order moments.

Second, we assessed convergence through ensemble consistency. We divided each acquired record into 10 independent segments, each corresponding to approximately $10^3$ integral time scales, and computed the turbulence statistics separately for each segment. Figure~\ref{fig:Figure3}(\textit{b}) shows the resulting estimates of $\overline{u'^2}$ for the individual segments. Most segment-wise estimates lie within the 95\% confidence interval of the mean obtained using the full acquisition record, indicating that the statistics remain converged across independent subsets of the data.

Finally, we examined the convergence of the tail of the probability distribution by evaluating the second-order integrand of the longitudinal velocity fluctuations, $u'^2 p(u')$,
where $p(u')$ denotes the probability density function of $u'$. Figure~\ref{fig:Figure3}(\textit{c}) shows that the integrand rises to the peak smoothly and decays rapidly to zero for large fluctuation amplitudes, indicating adequate sampling of rare events contributing to the second-order statistics. We observed similar convergence behavior for the remaining turbulent quantities considered in the present study, although we do not show those results here for brevity.

\subsubsection{Effectiveness of turning vanes in generating skewed mean flow}
\label{sec:RoleofTurningVanes}

Guide vanes have long been employed to manipulate and redirect flow in a wide range of engineering applications, including turbomachinery~\citep{korpela2019principles}, wind-energy systems~\citep{govardhan2002effect}, and closed-circuit wind tunnels~\citep{mehta1979design}. 
In the present study, we use airfoil-shaped turning vanes to impose controlled skew simultaneously on high- and low-speed streams of turbulent mixing layers. 
We set the vane deflection angles to $\pm 10^\circ$ (see \S~\ref{sec:ExptSetup}) and quantified the resulting flow skew by measuring the mean spanwise velocity component, $W$, together with the local spanwise flow angle,
$\gamma=\tan^{-1}\left({W}/{U}\right)$.

For the canonical planar mixing-layer configuration, symmetry requires $W=0$ everywhere, apart from small experimental uncertainties. 
In the present measurements, however, the planar configuration exhibited a residual spanwise velocity of up to approximately $5\%$ of the local mean streamwise velocity. 
This residual most likely arose from a slight misalignment of the measurement frame, possibly associated with X-wire misalignment or angular sensitivity during calibration \citep{Bruun1995}.
To remove this systematic bias, we corrected the velocity components by rotating the reference frame using the local flow angle measured in the nominally planar configuration at each downstream and cross-stream location. 
We then applied the same correction procedure consistently to both the planar and skewed datasets. This correction recovered the nominal planar configuration in the planar case and shifted the freestream flow angles in the skewed case closer to zero as expected, although not identically zero at all locations due to physical reasons discussed below. 
We verified that this procedure does not modify any of the principal trends, quantitative observations, or conclusions presented in this study.

\begin{figure}
 \centering
 \includegraphics[width=1\linewidth,trim={20 30 20 55},clip]{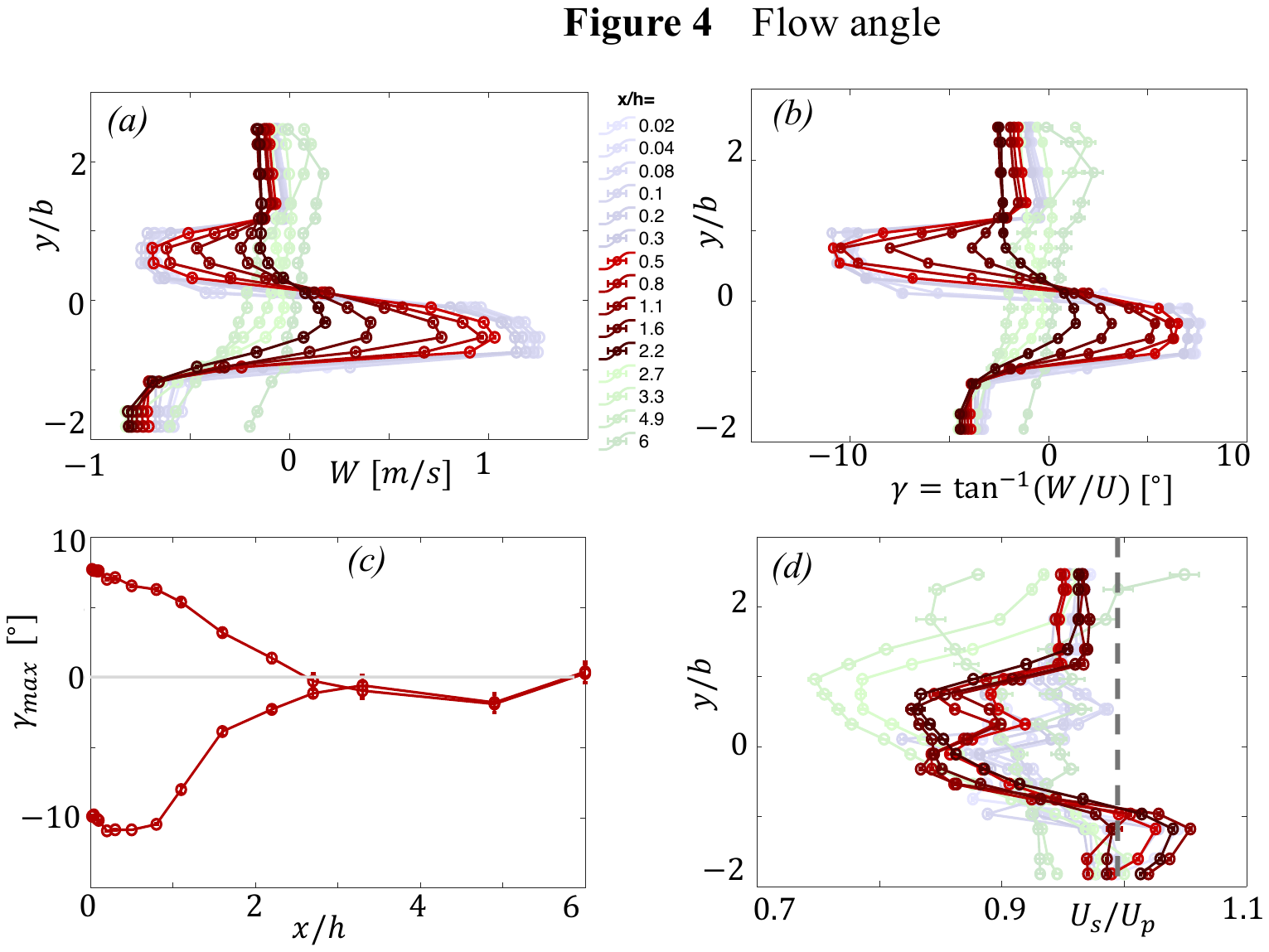}
\caption{Effectiveness of turning vanes in a generating skewed mean flow. 
\textit{(a)} Mean spanwise velocity and 
\textit{(b)} mean spanwise flow angles at various normalized vertical locations $y/b$ for several downstream locations, $x/h$, where $h$ and $b$ are the wind tunnel height and vane height respectively. 
$W$- and $U$ are the mean spanwise- and mean streamwise velocities, respectively. 
Colors correspond to different downstream locations. 
\textit{(c)} The maximum flow angle as a function of downstream distance. 
The flow angle decays to $1^\circ \pm 1^\circ$ at $x/h=2.2$. 
\textit{(d)} Ratio of skewed to planar mean velocities; $U_s$ and $U_p$ denote the mean streamwise velocities in the laboratory frame for skewed and planar mixing layers, respectively. 
The dashed grey line indicates $U_s/U_p = 1$. }
 \label{fig:Figure4}
\end{figure}

Figure~\ref{fig:Figure4}(\textit{a}) shows the measured profiles of $W$ for the skewed configuration at all downstream stations. 
The most significant observation is that the imposed turning generates a mean spanwise velocity across the mixing region, thereby confirming the effectiveness of the vane arrangement in producing a controlled three-dimensional mean flow. 
The skewing attains maximum value at $|y/b|\approx 0.5$, where $b$ is vane height, and remains primarily effective within the vane height, approximately corresponding to $|y/b|<1$. 
Outside this region, $W$ approaches zero on the low-speed side at all downstream locations. 
On the high-speed side, however, $W$ remains nonzero, with values of approximately $-0.7 \pm 0.2~\mathrm{m/s}$, and decays toward zero only at the furthest downstream station, $x/h=6$.

Because experimental constraints prevented measurements beyond approximately $y/b\approx -2$, we could not directly characterize the full cross-stream distribution of the flow. 
Nevertheless, conservation of spanwise mass flux,
\[
\int W\,dy \approx 0,
\]
implies that the negative spanwise velocity, with characteristic magnitude $W\approx -0.7~\mathrm{m/s}$, must persist further into the high-speed side of the tunnel cross-section.  The sustained non-zero values of $W$ outside the nominal vane region suggest the presence of large-scale circulation generated by the interaction of the skewed flow with the tunnel side walls. In particular, the negative sign of $W$ is consistent with a return-flow mechanism in which the laterally deflected flow impinges on the side wall, reverses direction, and subsequently recirculates across the tunnel cross-section while gradually losing momentum downstream.

Because we imposed equal but opposite vane deflections on either side of the splitter plate, the magnitude of $W$ within the vane region differs between the two streams, with the high-speed side exhibiting larger spanwise velocities than the low-speed side. 
To account for this effect, we examined the local flow angle $\gamma$ rather than the dimensional spanwise velocity itself. 
Figure~\ref{fig:Figure4}(\textit{b}) shows the downstream evolution of the local flow-angle profiles. When normalized in this manner, the maximum flow angles on the high- and low-speed sides become comparable and remain close to the imposed vane deflection angles, as also illustrated in figure~\ref{fig:Figure4}(\textit{c}). 

As the flow evolves downstream, the imposed skew progressively relaxes and the spanwise flow angle decreases monotonically with streamwise distance. 
By approximately $x/h = 2.2$, the peak flow angle reduces to values near $|\gamma| \approx 1^\circ \pm 1^\circ$, beyond which the flow approaches a nominally planar state. 
We therefore identify this location as an approximate upper limit of the downstream region over which the imposed mean flow three-dimensionality remains dynamically important. 
Interestingly, within the near field ($x/h \lesssim 1$), the high-speed stream exhibits turning angles slightly smaller than the imposed vane deflection. 
The physical mechanism responsible for this behavior remains unclear at the time of writing.

To quantify the influence of skewing on the streamwise velocity magnitude, we examined the ratio between the mean velocity of the skewed case, $U_s$, and that of the planar case, $U_p$ (figure~\ref{fig:Figure4}(\textit{d})). 
The ratio varies approximately within the range $0.80 \lesssim U_s/U_p \lesssim 1.05$. 
At most downstream and cross-stream locations, the skewed configuration exhibits lower streamwise velocities than the planar case. 
The largest reductions occur within the vane region, where the imposed spanwise turning remains strongest (see figure~\ref{fig:Figure4}(\textit{b}). 
The maximum reduction in the streamwise velocity component, $(1-U_s/U_p)$, approximately matches the magnitude of the imposed vane deflection $(=\tan10^\circ)$. 
Outside the vane region, however, the ratio approaches unity, and in portions of the high-speed side for $0.5 \lesssim x/h\lesssim 2$, the skewed case even exhibits slightly larger streamwise velocities than the planar configuration.

Overall, the turning-vane arrangement provides a simple and experimentally realizable method to generate controlled mean-flow skew in turbulent mixing layers. 
The close agreement between the imposed vane deflection angles and the measured spanwise flow angles, together with the corresponding reduction in the streamwise velocities, demonstrates that the apparatus effectively generates the intended mean-flow three-dimensionality and provides a practical way to systematically investigate its influence on the development of turbulent free-shear flows.

\subsubsection{Region of interest}
\label{sec:RegionofInterest}

Although the present configuration substantially simplifies the generation of skewed mixing layers by using a single-contraction facility, the introduction of a splitter plate, turning vanes, and vane end plates inevitably introduces additional geometric complexity into the flow. 
These elements generate localized disturbances and boundary-condition effects that can influence the near-field evolution of the mixing layer independently of the imposed mean-flow three-dimensionality itself. 
It is therefore necessary to distinguish between flow features arising directly from the experimental apparatus and those associated with the intrinsic dynamics of the skewed mean flow. 
To achieve this separation, we identify and focus on an intermediate downstream region in which the influence of the immediate geometric forcing has sufficiently weakened while the imposed mean flow three-dimensionality remains dynamically significant. 
Hereinafter, we refer to this region as the \textit{region of interest}.

\begin{figure}
 \centering
 \includegraphics[width=0.9\linewidth, trim={20 200 20 100},clip]{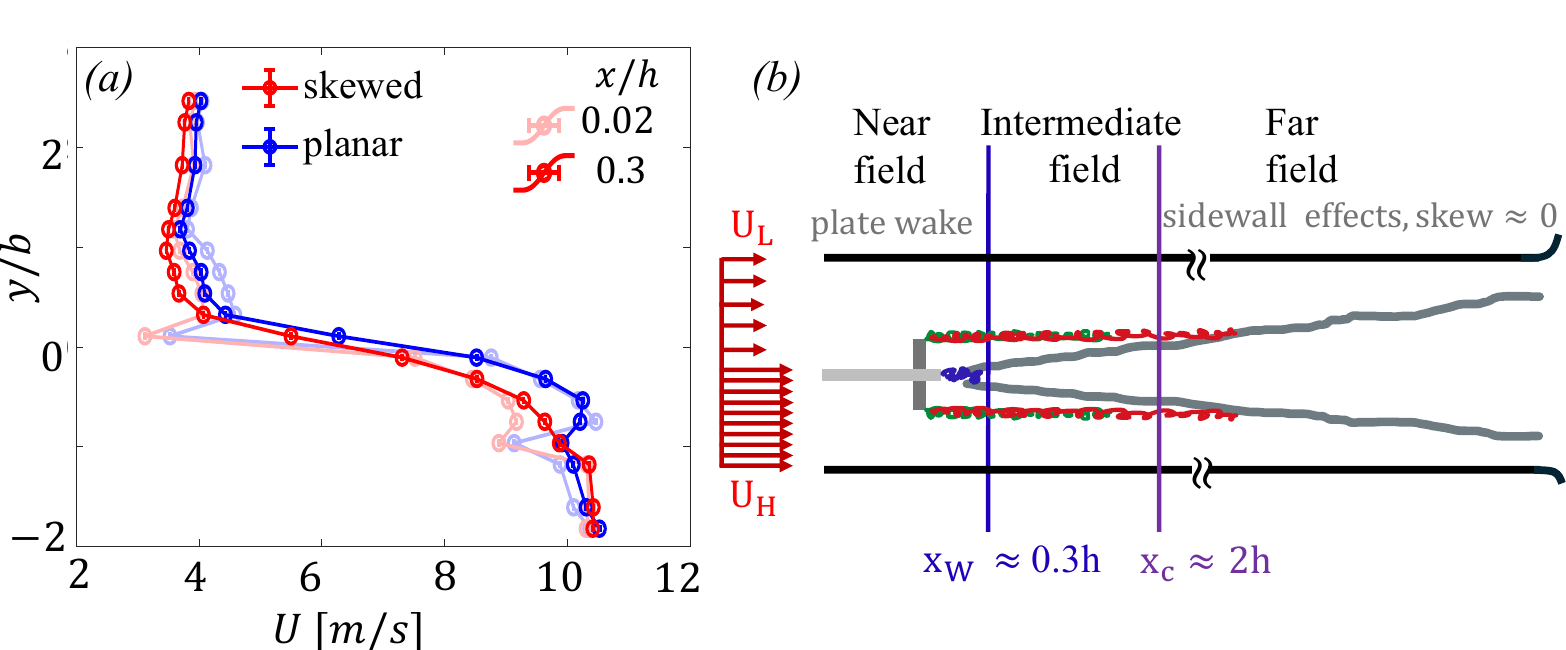}
\caption{\textit{(a)} Mean streamwise velocities at x/h =0.02 and 0.3 for skewed (Red) and planar (blue) cases, respectively. 
The velocity deficit near $y/b \approx 0$ is likely associated with the splitter-plate wake, whereas the deficits near $|y/b| \approx 1$ arise from the end-plate wakes. 
\textit{(b)} Schematic showing the region of interest between initial times where the plate wake is pronounced and later times where side wall effects are pronounced. 
Plate wakes are shown in blue scribbles. 
Also shown here are end-plate wakes (green), and the streamwise vorticity (red) generated by misalignment between the inner skewed mixing region and the outer unskewed freestream. $b$ and $t$ are vane height and splitter plate thickness, respectively. }
 \label{fig:Figure5}
\end{figure}

As described in \S~\ref{sec:ExptSetup}, the high- and low-speed streams are separated by a splitter plate with a blunt trailing edge of thickness $t \approx 1.2~\mathrm{cm}$. 
The nonzero thickness of the splitter plate generates a wake downstream of the trailing edge, which manifests as a streamwise velocity deficit centered near the mixing-layer interface, as shown in figure~\ref{fig:Figure5}(\textit{a}).
We observe the velocity defect primarily in the vicinity of $y/b=0$, and it persists up to approximately $x_w/h \approx 0.3$, corresponding to $x_w/t \approx 30$. 
This extent agrees well with previous investigations of splitter-plate wakes in turbulent mixing layers~\citep{laizet2010direct,braud2004low}. 
We therefore designate the region $x_w/h < 0.3$ as the \textit{wake-dominated} or \textit{initial-transient} regime (figure~\ref{fig:Figure5}(\textit{b})). 

Further downstream, turbulent mixing layers may eventually evolve toward an asymptotic state~\citep{Pope_2000}. 
However, the downstream distance required to attain such a state depends sensitively on the inflow conditions and boundary constraints, including the characteristics of the merging boundary layers~\citep{bell1990development}, incoming turbulence intensity~\citep{chandrsuda1978effect}, externally imposed perturbations~\citep{nygaard1991evolution}, splitter-plate geometry~\citep{mehta1991effect}, and finite-domain effects~\citep{mcmullan2015spanwise}. 
In the present configuration, side-wall confinement becomes particularly important because the imposed skew introduces a spanwise velocity, as discussed in \S~\ref{sec:RoleofTurningVanes}. 
A fluid particle starting at the center of the splitter plate trailing edge would reach the side wall at $x/h \sim (U_L+U_H)/(4W) \approx 2$. 
We should expect spanwise confinement effects to start before this point, and to be strong after this point. 
Consistent with this estimate, the spanwise flow angle decays progressively downstream and approaches values close to zero beyond approximately $x_c/h \gtrsim 2$ (see \S~\ref{sec:RoleofTurningVanes}). 
This observation suggests that the imposed skew loses much of its dynamical influence beyond this location, in agreement with previous numerical observations indicating that skew-induced effects primarily influence the near- and intermediate-field development of the mixing layer~\citep{meldi2020numerical}.

To minimize contamination from splitter-plate wake effects, side-wall confinement, while retaining the dominant influence of the imposed mean-flow skew, we restrict the present analysis to the intermediate downstream region $0.5 \le x/h \lesssim 2$, our \textit{region of interest}. 
In terms of the initial momentum thickness, this region corresponds approximately to $40$--$180$ times the combined initial momentum thickness of the two incoming streams.

\subsection{Companion LES}
\label{sec:LES}

The companion large-eddy simulation (LES) study \citep{kumar2024three} considers an idealized turbulent mixing layer in a temporally evolving framework with periodic boundary conditions in the splitter-plate-parallel directions. The simulations therefore eliminate spanwise confinement effects and do not model the finite test-section geometry of the experiments.
The LES initial condition is constructed from two independently developed turbulent boundary layers, at which point they are stitched together at their interface to form the initial mixing layer. This procedure represents an idealized flow state immediately downstream of the splitter-plate trailing edge and neglects the trailing-edge wake. 
While this assumption is not realistic in the immediate vicinity of the splitter plate, it is known to produce accurate shear layer dynamics farther downstream~\citep{Sandham_JT_2009, pirozzoli2015early,laizet2010direct}.
In addition, the skewed case has a different angle between the
streams in the LES compared to the experiments (when measured for the initial boundary layer states in the LES).

Due to all these factors, quantitative agreement is not expected between the LES and the experiments, particularly during the early stages of development when the flow remains strongly influenced by its initial conditions. Nevertheless, comparison between the two datasets provides a valuable consistency check and helps identify flow features that are robust to the details of the configuration.  In the following, we therefore compare the experimentally measured statistics with the LES results at three representative stages of temporal evolution corresponding to early, intermediate, and long times. It is noted that the companion LES and literature data have been appropriately rotated to ensure consistency with the present coordinate system and problem definition.

\section{Results}
\label{sec:3}

In the following sections, we examine the mean-flow evolution and turbulence statistics of both the planar and skewed mixing layers. 
Unless stated otherwise, we restrict the discussion to the region of interest identified in \S~\ref{sec:RegionofInterest}, where the flow evolution is minimally influenced by splitter-plate wake effects and side-wall confinement, while the imposed mean-flow three-dimensionality remains dynamically significant. 
In \S~\ref{sec:MeanFlowQuantities}, we analyze the mean-flow properties, including the streamwise velocity distribution, mixing-layer thickness, and mean vorticity field. 
We then examine the evolution of the turbulent stresses and kinetic energy in \S~\ref{sec:turbulentquantities}.

\subsection{Mean flow quantities}\label{sec:MeanFlowQuantities}

\subsubsection{Mean streamwise velocity}\label{sec:MeanVelocity}

\begin{figure}
 \centering
 \includegraphics[width=1\linewidth,trim={40 280 50 60},clip]{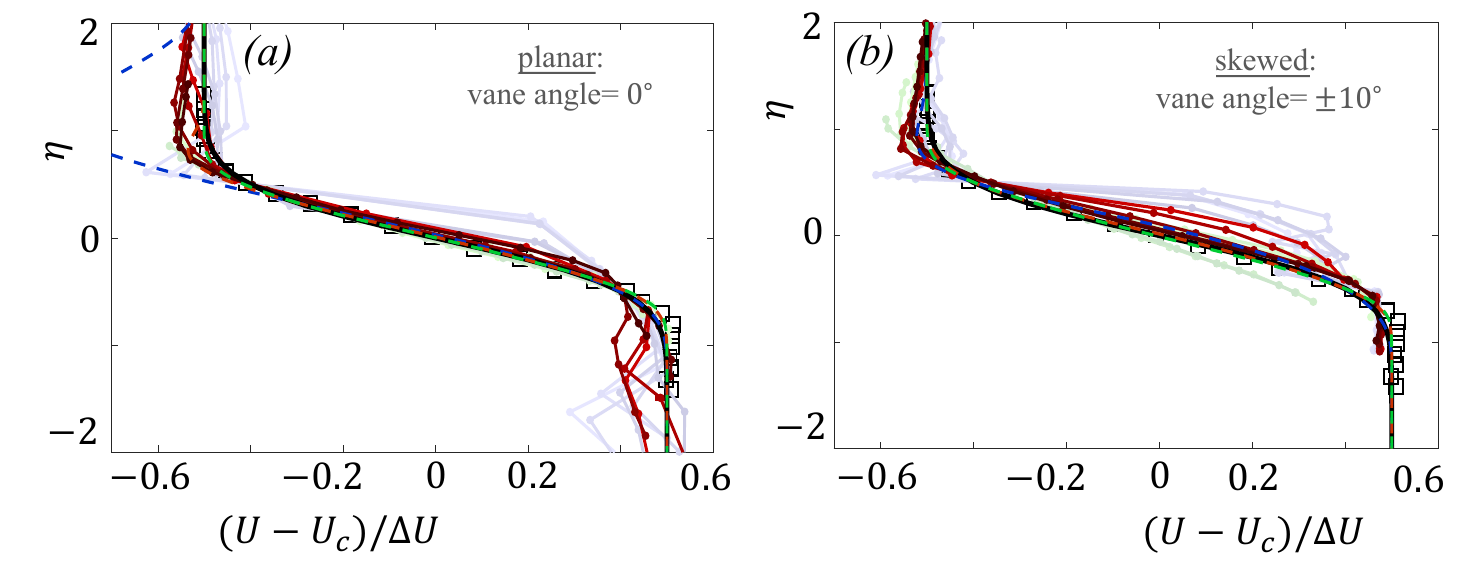}
\caption{Mean streamwise velocity profiles in a moving reference frame at various downstream locations. 
\textit{(a)} Planar mixing layers. 
\textit{(b)} Skewed mixing layers. 
Colored circles correspond varying downstream distance in the present experiments. 
Legends are as in figure~\ref{fig:Figure4}. 
The dashed coloured lines represent the companion LES at different time instances: blue (initial), red (intermediate), and green (late time). 
The black line is an analytical model for the shape of the mean velocity profile~\eqref{eqn:erfVel}. 
The square symbols denote the experimental data in \citet{bell1990development} for planar mixing layers. 
The LES data are configuration-specific, whereas the square symbols shown in (a) and (b) correspond to the planar case only. 
The companion LES and reference data have been appropriately rotated to ensure consistency with the present coordinate system and problem definition. 
Here, $U_c = (U_H + U_L)/2$ is the mean convection velocity and $\Delta U = U_H - U_L$ is the velocity difference across the shear layer. 
The cross-stream coordinate is expressed in terms of the similarity variable $\eta$ (see main text for definition). 
The mean profiles exhibit approximate collapse when scaled with $-(1/2) \mathrm{erf}(\eta/0.55)$ for both configurations. }
 \label{fig:Figure6}
\end{figure}

Figure~\ref{fig:Figure6} shows the mean streamwise velocity profiles for the planar and skewed mixing layers at different downstream locations. Following the conventional normalization used for turbulent mixing layers~\citep{Pope_2000}, we express the velocity in a convecting reference frame and normalize it using the freestream velocity difference,
$f(\eta)={(U-U_c)}/{\Delta U}$,
where $\Delta U = U_H-U_L$ and $U_c={(U_H+U_L)}/{2}$.
We similarly define the cross-stream coordinate using the similarity variable $\eta={(y-y_{\mathrm{avg}}(x))}/{\delta_U(x)}$,
where $ y_{avg}(x)= (y(U=0.9\Delta U)+y(U=0.1\Delta U))/2, \delta_U(x)= y(U=0.1\Delta U)-y(U=0.9\Delta U))$. Negative values of $\eta$ correspond to the high-speed side of the mixing layers, whereas positive values correspond to the low-speed side.

For both planar and skewed configurations, the mean velocity approaches the corresponding freestream values for $|\eta|\gtrsim1$ at all measured downstream stations (figure~\ref{fig:Figure6}(\textit{a,b})). 
When expressed in similarity coordinates and in the convective frame, the profiles collapse, indicating an approximately self-preserving evolution within the region of interest. Similar behavior is observed in the companion LES and in the experimental study of \citet{bell1990development}. 
The onset and extent of this collapse, however, differs between the planar and skewed cases. 
For the planar mixing layer, the profiles collapse reasonably well for all measured downstream locations satisfying $x/h \ge 0.5$. 
In contrast, the skewed mixing layer exhibits comparable collapse only over the more limited range $1\lesssim x/h \lesssim 3$. 
\citet{azim2003plane} reports similar behavior for skewed mixing layers with imposed skew angles of $18^\circ$, with a profile collapse that persisted further downstream than we observed in our experiments. 
In the present skewed configuration, the breakdown of collapse beyond approximately $x/h\gtrsim3$ likely reflects the increasing influence of side-wall confinement, as discussed in \S~\ref{sec:RegionofInterest}. 

The collapsed profiles are well approximated by the error-function~\citep{Pope_2000}

\begin{equation} 
\frac{U-U_c}{\Delta U}
=
f(\eta)
=
-\frac{1}{2}\,
\mathrm{erf}
\left(
\frac{\eta}{\sigma\sqrt{2}}
\right),
\qquad
\sigma=
\left[
2\sqrt{2}\,
\mathrm{erf}^{-1}\left(\frac{4}{5}\right)
\right]^{-1}.
\label{eqn:erfVel}
\end{equation}
In the near field ($x/h<0.3$), the low-speed side exhibits a pronounced velocity deficit within the interval $0<\eta<1$. 
As discussed in \S~\ref{sec:RegionofInterest}, this deficit originates from the splitter-plate wake generated by the blunt trailing edge. 
Notably, the wake remains preferentially confined to the low-speed side of the mixing layer. 
Previous studies have reported similar asymmetric wake localization in turbulent mixing layers generated with splitter plates~\citep{braud2004low,azim2003plane,mehta1991effect}. 
A possible explanation follows from the vortex-shedding dynamics associated with asymmetric freestream velocities. 
\citet{boldman1976vortex} investigates vortex shedding behind blunt trailing edges of thickness comparable to that used here and shows that, for velocity ratios $r<1$, vortices shed into the high-speed stream diffuse more rapidly, whereas those on the low-speed side remain weaker and more spatially localized. 
The stronger wake deficit observed in the skewed configuration suggests that the imposed mean flow three-dimensionality amplifies the persistence of the wake-induced velocity defect.

We also observe localized streamwise velocity deficits near the vane height for both planar and skewed configurations within the near field ($x/h\lesssim1.5$). 
This feature appears particularly clearly in figure~\ref{fig:Figure5}(\textit{a}) when the cross-stream coordinate is normalized using the vane height. 
In the planar configuration, the velocity decreases by as much as approximately $25\%$ of $\Delta U$ on the high-speed side, whereas the corresponding reduction on the low-speed side remains below approximately $5\%$ of $\Delta U$. 
The skewed configuration exhibits slightly larger deficits, reaching approximately $30\%$ and $10\%$ of $\Delta U$ on the high- and low-speed sides, respectively. 
The velocity-deficit region also remains substantially narrower on the high-speed side, likely because stronger streamwise convection suppresses lateral momentum diffusion. 
The confinement of these deficits near the vane height suggests that they originate from the wake generated by the vane end plates rather than from the mixing-layer dynamics themselves.

\subsubsection{Shear thickness}\label{sec:ShearThickness}

\begin{figure}
 \centering
 \includegraphics[width=1\linewidth,trim={20 200 20 98},clip]{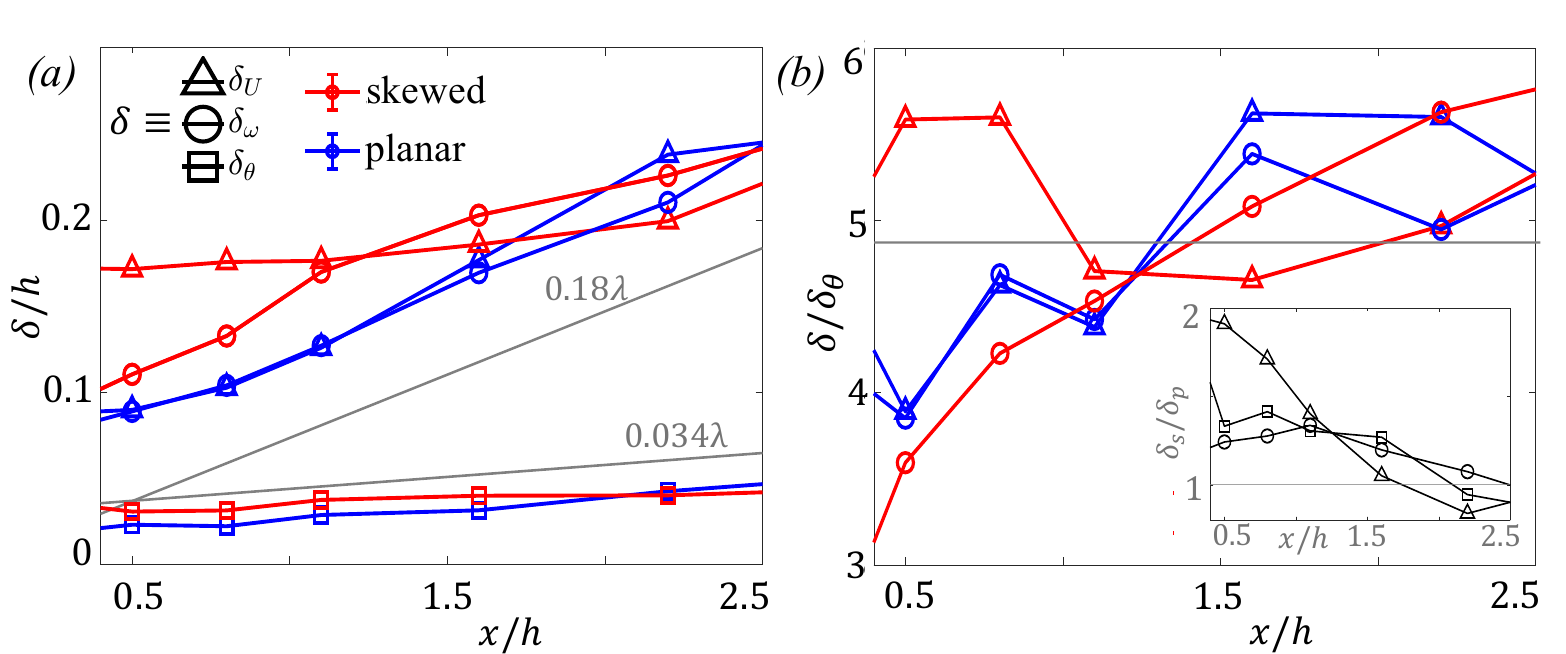}
\caption{\textit{(a)} Streamwise development of different mixing layer thicknesses, $\delta/h$ for skewed (red) and planar (blue) cases. Grey line: upper $d\delta/dx=0.18\lambda$ and lower $d\delta/dx=0.034\lambda$. 
\textit{(b)} $\delta_U$ and $\delta_{\omega}$ with respect to $\delta_{\theta}$. 
Inset shows the ratio of skewed mixing thicknesses ($\delta_s)$ to the planar ones ($\delta_p$). Symbols as in \textit{(a)}. 
The grey line is $\delta/\delta_{\theta}=4.8$. }
 \label{fig:Figure7}
\end{figure}

Figure~\ref{fig:Figure7} shows the downstream evolution of the mixing-layer thicknesses for both planar and skewed configurations within the region of interest. 
Previous studies have used several definitions to characterize the growth of turbulent mixing layers. 
In the present work we use three standard measures, which were based on either the mean velocity profile, the mean shear, or the momentum deficit. 

First, following the definition introduced in \S~\ref{sec:MeanVelocity}, we define the velocity thickness as~\citep{Pope_2000}

\begin{equation}
\delta_U= y_{(U=0.1\Delta U)}-y_{(U=0.9\Delta U)}.
\label{eqn:delta_U}
\end{equation}

Second, we define the vorticity thickness using the maximum mean velocity gradient~\citep{brown1974density},

\begin{equation}
\delta_{\omega}=\frac{\Delta U}{\left(\partial U/\partial y\right)_{\max}}.
\label{eqn:delta_omega}
\end{equation}

Finally, we define the momentum thickness as~\citep{rogers1994direct}

\begin{equation}
\delta_{\theta}=\int_{-\infty}^{\infty}\left[\frac{1}{4}-
\left(\frac{U-U_c}{\Delta U}\right)^2\right]dy.
\label{eqn:delta_theta}
\end{equation}

For self-similar planar turbulent mixing layers, classical analyses by \citet{abramovivc1963theory} and \citet{sabin1965analytical} suggested that the spreading rate scales linearly with the velocity-ratio parameter 
$\lambda=\frac{1-r}{1+r}$,
such that

\begin{equation}
\frac{d\delta}{dx}
=
c_{\delta}\lambda,
\label{eqn:sabin}
\end{equation}

where $r=U_L/U_H$. 
For the vorticity thickness, previous experimental investigations have reported values in the range $0.125 \lesssim c_{\delta_\omega}\lesssim 0.225$~\citep{dimotakis1989turbulent}. 

Empirical studies further suggest approximate proportionality between the spreading rates defined using different thickness measures~\citep{brown1974density,browand1985turbulent},

\begin{equation}
\frac{d\delta_\omega/dx}{0.181}
=
\frac{d\delta_\theta/dx}{0.034}
=
\lambda.
\label{eqn:delta_BrownRoshko}
\end{equation}

More recently, \citet{yoder2015modeling} shows that, for a mean velocity profile approximated by the error-function form originally proposed in \citet{gortler1942berechnung}, the spreading rates satisfy

\begin{equation}
\frac{d\delta_U/dx}{0.165}
=
\frac{d\delta_\omega/dx}{0.161}
=
\frac{d\delta_\theta/dx}{0.036}
=
\lambda.
\label{eqn:delta_Yoder}
\end{equation}

As shown in figure~\ref{fig:Figure7}(\textit{a}), all three thickness measures, $\delta_U$, $\delta_\omega$, and $\delta_\theta$, remain consistently larger in the skewed mixing layer than in the planar case throughout the region of interest, $0.5\le x/h\lesssim2$. 
This behavior agrees qualitatively with previous observations in both compressible~\citep{boukharfane2021direct} and incompressible~\citep{azim2003plane} skewed mixing layers. 
The enhanced thickness growth in the skewed configuration likely reflects the additional lateral transport and deformation introduced by the imposed mean-flow three-dimensionality, particularly in the near field where the spanwise flow angles remain large (see figure~\ref{fig:Figure4}). 
Beyond approximately $x/h\gtrsim2$, however, the skewed thicknesses become smaller than their planar counterparts (see figure~\ref{fig:Figure7}(\textit{b inset}). 
As discussed previously, this transition likely results from near-zero flow turning, increasing side-wall confinement and finite-domain effects, which progressively modify the downstream evolution of the skewed flow.

Both the planar and skewed mixing layers exhibit approximately linear growth of the vorticity thickness within the region of interest, with ${d\delta_\omega}/{dx}\approx0.075$. 
Expressed in the form of \eqref{eqn:sabin}, this corresponds to
$c_{\delta_\omega}\approx0.17$, which agrees closely with the canonical measurements in \citet{brown1974density} and lies within the range summarized in \citet{dimotakis1989turbulent}. 
Similarly, the momentum thickness $\delta_\theta$ also grows approximately linearly with downstream distance for both configurations, at approximately similar rate, $c_{\delta_\theta}\approx0.03$, 
consistent with the proportionalities given by \eqref{eqn:delta_BrownRoshko} and \eqref{eqn:delta_Yoder}.

The evolution of the velocity thickness, $\delta_U$, differs from that of the vorticity thickness, $\delta_\omega$, and exhibits a notable dependence on the imposed mean-flow three-dimensionality. 
For the planar configuration, $\delta_U$ increases approximately linearly with downstream distance and remains comparable in magnitude to $\delta_\omega$ throughout the region of interest. 
In contrast, the skewed mixing layer exhibits a substantially weaker growth of $\delta_U$. 
The velocity thickness is initially larger than $\delta_\omega$ and subsequently evolves more slowly downstream.

Discrepancies among different measures of the mixing-layer thickness during the early stages of development have been reported previously and are often attributed to transient effects associated with the approach toward a self-preserving state \citep{Pope_2000,pirozzoli2015early}. 
Reduced growth rates of $\delta_U$ relative to other characteristic thicknesses have also been reported in compressible planar mixing layers, where they are commonly associated with suppressed mixing \citep{gatski2013compressibility}. 
While the present flow is incompressible and the underlying mechanism is therefore unlikely to be the same, these observations highlight that different thickness measures need not evolve identically during non-equilibrium stages of development. 
The physical origin of the reduced growth of $\delta_U$ in the present skewed configuration remains unclear. 
At present, the results indicate only that the imposed mean-flow three-dimensionality influences the evolution of the mean velocity profile and the mean shear differently during the transient development of the mixing layer.

Least-squares fits to the measured thickness evolution yield the following spreading-rate relations for the planar mixing layer:

\begin{equation}
\frac{d\delta_U/dx}{0.178}
=
\frac{d\delta_\omega/dx}{0.167}
=
\frac{d\delta_\theta/dx}{0.03}
=
\lambda.
\nonumber
\end{equation}

For the skewed mixing layer, the least-squares fits yield

\begin{equation}
\frac{d\delta_U/dx}{0.04}
=
\frac{d\delta_\omega/dx}{0.158}
=
\frac{d\delta_\theta/dx}{0.026}
=
\lambda.
\nonumber
\end{equation}

Figure~\ref{fig:Figure7}(\textit{b}) shows the ratios of the different thickness measures normalized by the momentum thickness. 
For the planar configuration, both $\delta_\omega/\delta_\theta$ and $\delta_U/\delta_\theta$ vary between approximately 4 and 6 with a mean value close to 4.8. 
These values agree well with previous measurements reported in \citet{li2010experimental} and \citet{rogers1994direct}. 
\citet{rogers1994direct} notes that these ratios depend sensitively on the detailed shape of the mean velocity profile. 
For example, an error-function profile yields $\delta_\omega/\delta_\theta\approx4.44$, whereas a hyperbolic-tangent profile yields a value closer to 4. 
The present skewed mixing layer exhibits similar behavior, at least for the ratio $\delta_U/\delta_\theta$.

\subsubsection{Mean velocity gradients and vorticity}
\label{sec:velGrad}

The velocity difference across the interface between the high- and low-speed streams generates a vortex sheet, which when acted upon by Kelvin--Helmholtz instability, redistributes momentum and progressively smoothens the velocity discontinuity~\citep{Pope_2000}. 
This mechanism operates in both planar and skewed mixing layers. 
To quantify the resulting mean vorticity field, we computed the local spatial velocity gradients using finite-difference approximations. 
We applied a second-order central-difference scheme to the interior measurement locations within the core region of the flow and used first-order forward or backward differences near the measurement boundaries.

\begin{figure}
 \centering
 \includegraphics[width=1\linewidth,trim={10 20 50 50},clip]{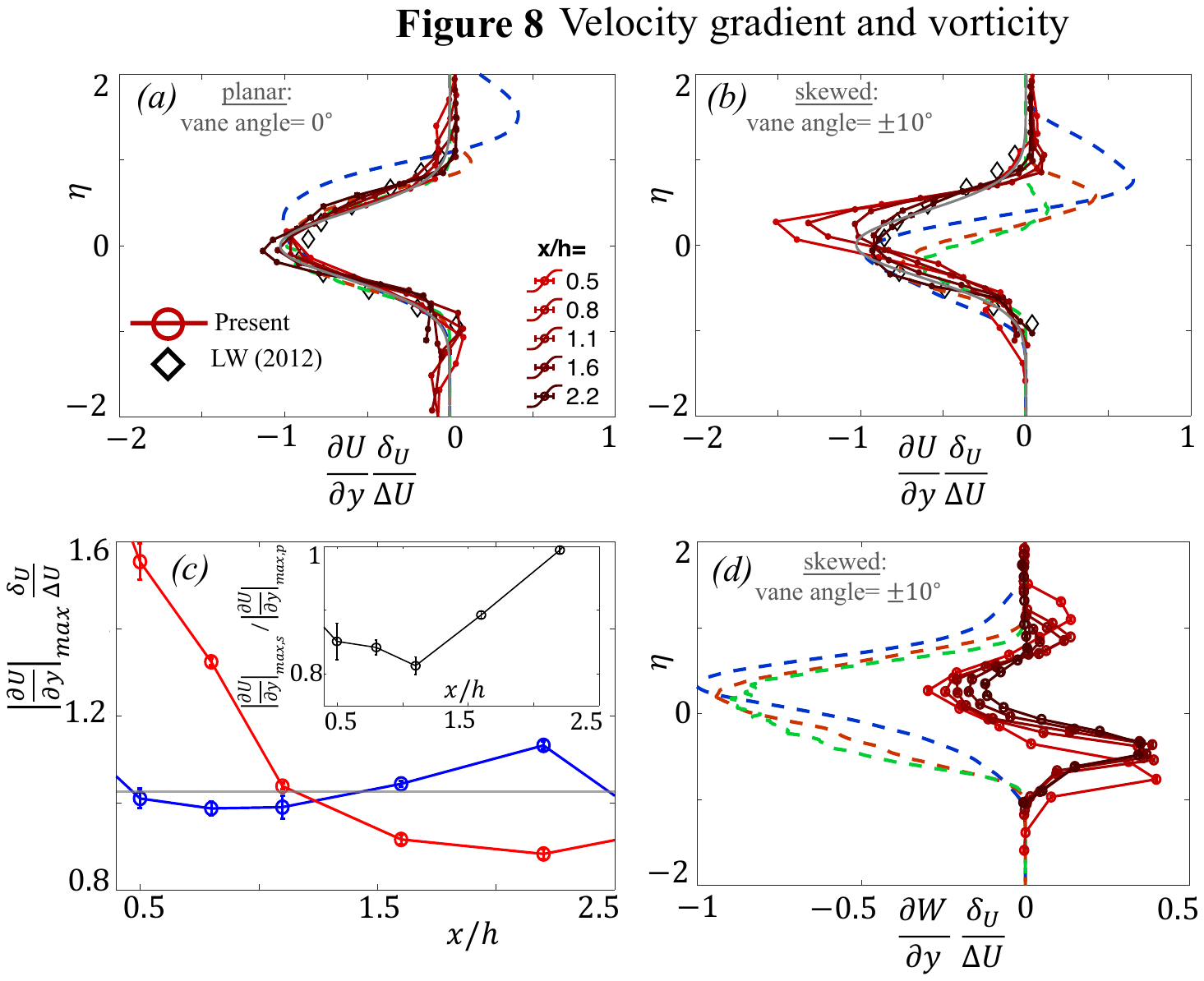}
\caption{ Normalized spanwise vorticity profiles for planar 
\textit{(a)} and skewed cases 
\textit{(b)} in region of interest; different shades of red correspond to different downstream locations in the present experiments, consistent with that used in figure.~\ref{fig:Figure4}.
The dashed coloured lines represent the companion LES at different time instances: blue (initial), red (intermediate), and green (late time). 
The grey line is the analytical model obtained by taking the derivative of the mean velocity profile~\eqref{eqn:velGrad}. 
The diamond symbols are from \citet{loucks2012velocity}. 
The LES data are configuration-specific, whereas the diamond symbols shown in (a) and (b) correspond to the planar case only. 
\textit{(c)} streamwise development of the normalized maximum values of the spanwise vorticity. 
Red and blue correspond to the skewed and planar cases, respectively. 
The gray horizontal line is the maximum of an error-function fit to the velocity profile \eqref{eqn:velGrad}. 
The inset shows the ratio of the skewed maximum velocity gradient to the planar one. 
\textit{(d)} Normalized streamwise vorticity profiles for skewed mixing layers. 
Legend as in \textit{(a)} and \textit{(b)}. }
 \label{fig:Figure8}
\end{figure}

For planar mixing layers, the dominant mean vorticity component is the spanwise vorticity, $\omega_z={\partial V}/{\partial x}-{\partial U}/{\partial y}$. 
Because the planar configuration is statistically homogeneous in the spanwise direction and the mean cross-stream velocity remains small \citep{bell1990development,rogers1994direct}, the dominant contribution reduces to approximately
$\omega_z \approx -{\partial U}/{\partial y}$.
Figures~\ref{fig:Figure8}(\textit{a,b}) show the measured spanwise vorticity distributions for the planar and skewed configurations, respectively. 
We compare the measured profiles of the normalized velocity gradient with the analytical distribution obtained by differentiating the error-function representation of the mean velocity profile introduced in \S~\ref{sec:MeanVelocity}. 
Differentiating \eqref{eqn:erfVel} yields

\begin{equation}
\frac{\partial U}{\partial y}
=
-\frac{\Delta U}{\sigma\sqrt{2\pi}}
\exp
\left[
-
\left(
\frac{\eta}{\sigma\sqrt{2}}
\right)^2
\right]
\frac{\partial\eta}{\partial y}.
\label{eqn:velGrad}
\end{equation}

For the planar mixing layer, the normalized spanwise-vorticity profiles remain approximately symmetric about the centerline and agree reasonably well with the analytical prediction, the companion LES and the experimental measurements in \citet{loucks2012velocity}(figure~\ref{fig:Figure8}(\textit{a})). 
In contrast, the skewed mixing layer exhibits a clear asymmetry, with the vorticity distribution biased toward the low-speed side of the flow (figure~\ref{fig:Figure8}(\textit{b})). 
A weaker form of this asymmetry also appears in the planar configuration very near the splitter plate ($x/h=0.5$), but the profiles recover approximate symmetry further downstream by approximately $x/h=2.2$. 
Both configurations also exhibit a weak positive dip in $\omega_z$ near $|\eta|\approx1$. 

The peak normalized spanwise vorticity in the planar mixing layer fluctuates between approximately $0.9$ and $1.2$, with a mean value close to $1$, in excellent agreement with the analytical peak value predicted by \eqref{eqn:velGrad} within the region of interest (figure~\ref{fig:Figure8}(\textit{c})). 
In the skewed configuration, however, the peak normalized vorticity decreases monotonically with downstream distance. 
Similar to the behavior observed for the mixing-layer thicknesses, the skewed layer initially exhibits larger normalized vorticity magnitudes than the planar case for $x/h\lesssim2$, after which the trend reverses and the skewed case becomes comparatively weaker. 
Interestingly, this trend reverses when the dimensional vorticity magnitude is considered instead: the absolute peak values of $\omega_z$ remain systematically smaller in the skewed configuration throughout the measured region, as shown in the inset of figure~\ref{fig:Figure8}(\textit{c}). 
This behavior possibly reflects the combined influences of the larger velocity thickness and the altered mean-shear distribution produced by the imposed skew.

Unlike the planar case, the skewed mixing layer contains directional shear associated with the imposed mean spanwise velocity varying in the vertical direction (see \S~\ref{sec:RoleofTurningVanes}). 
In addition to the conventional spanwise vortex sheet generated by the streamwise velocity difference, the directional shear introduces additional streamwise-oriented vortex sheets associated with gradients of the spanwise velocity. 
Specifically, the imposed skew generates one streamwise vortex sheet near the primary mixing-layer interface and two additional sheets near the boundaries separating the skewed flow within the vane region from the outer unskewed freestream (see \S~\ref{sec:RoleofTurningVanes}). 
To characterize these structures, we computed the mean streamwise vorticity,
$\omega_x={\partial W}/{\partial y}-{\partial V}/{\partial z}$, at all measured cross-stream and downstream locations. 
Figure~\ref{fig:Figure8}(\textit{d}) shows that the normalized streamwise vorticity changes sign at the interface. 
The vorticity is predominantly negative within the central skewed region and positive near the vane boundaries on either side of the mixing layer. 
We associate the negative central region with the streamwise vortex sheet generated at the primary mixing interface, whereas the outer positive regions likely correspond to streamwise vorticity generated at the interfaces between the inner skewed flow and the outer unskewed freestream near the vane boundaries. This interpretation is further supported by the companion LES study, which reproduces the central negative streamwise-vorticity region within the mixing layer but does not exhibit the outer positive-vorticity regions. 
In the LES study, only the primary skewed interface is present, whereas the secondary interfaces between the skewed and unskewed outer streams are absent by construction.
The magnitude of $\omega_x$ near the vane boundaries is also systematically larger on the high-speed side of the flow, consistent with the larger spanwise velocities measured there (see \S~\ref{sec:RoleofTurningVanes}). 
These observations demonstrate that the imposed skew possibly modifies the mean rotational structure of the flow by introducing persistent streamwise-oriented mean vorticity in addition to the conventional spanwise shear layer. 

\subsection{Turbulent quantities}
\label{sec:turbulentquantities}
\subsubsection{Primary shear stress}
\label{sec:PrimaryShear}

\begin{figure}
 \centering
 \includegraphics[width=1\linewidth,trim={40 30 40 60},clip]{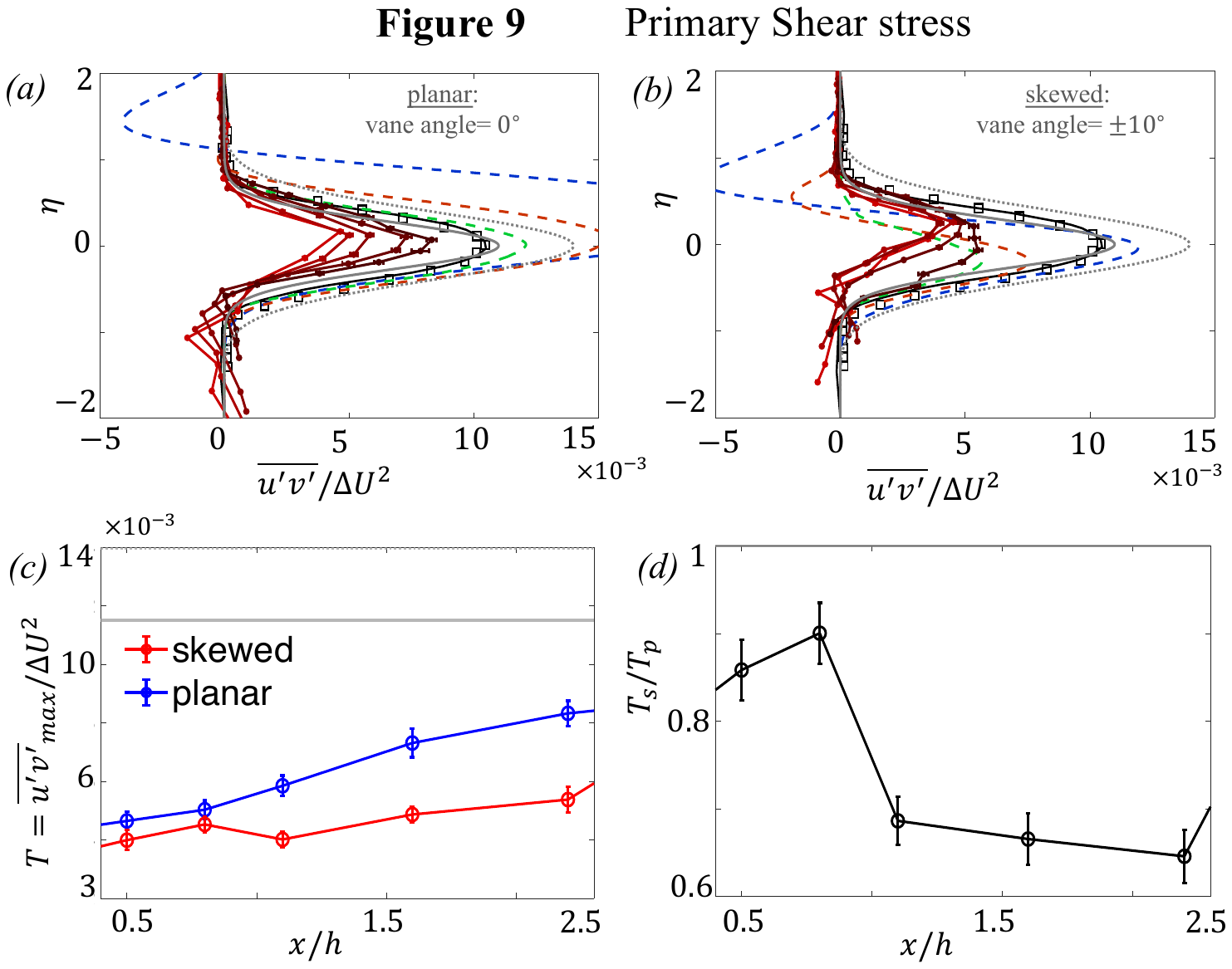}
\caption{ Profiles of primary Reynolds shear stress. 
\textit{(a)} Planar mixing layers. 
\textit{(b)} Skewed mixing layers. 
Red shades correspond to varying downstream distances in the present experiments, as shown in Fig. \ref{fig:Figure4}. 
The dashed coloured lines represent the companion LES at different time instances: blue (initial), red (intermediate), and green (late time). 
The square symbols denote the experimental data in \citet{bell1990development}, 
while the solid black line corresponds to the DNS results in \citet{rogers1994direct}. 
The companion LES results are configuration-specific, whereas the square symbols and black curve shown in \textit{(a)} and \textit{(b)} correspond to the planar case only. 
Also shown are asymptotic profiles obtained from \eqref{eqn:Restress} using the maximum value predicted by \eqref{eqn:maxRestressprofile} (grey dotted line) and an empirical maximum value of 0.011 (grey solid line). 
\textit{(c)} Streamwise evolution of the normalized maximum Reynolds shear stress $T$; 
blue and red markers correspond to the planar and skewed cases, respectively. 
The grey solid line indicates the maximum value reported in the literature \citep{yoder2015modeling}, while the grey dotted line corresponds to the theoretical maximum in the asymptotic state obtained from \eqref{eqn:maxRestressprofile} \citep{Pope_2000}. 
\textit{(d)} Ratio of the normalized maximum shear stress in skewed ($T_s$) to planar ($T_p$) mixing layers. }
 \label{fig:Figure9}

\end{figure}

Figure~\ref{fig:Figure9}(\textit{a,b}) shows the profiles of the normalized primary Reynolds shear stress, $\overline{u'v'}/\Delta U^2$, for both planar and skewed mixing layers at various downstream locations within the region of interest. 
Following standard convention, we normalize the stress using the square of the freestream velocity difference, $\Delta U^2$. 
In contrast to the mean velocity profiles, the normalized shear-stress distributions do not collapse within the core region of the mixing layer for either configuration. 
Outside the mixing region, however, the profiles exhibit partial collapse on the low-speed side. 
For the planar mixing layer, this collapse occurs for approximately $\eta \gtrsim 0.8$, whereas for the skewed case it begins further inward, at approximately $\eta \gtrsim 0.4$.

Both configurations exhibit weak negative values of $\overline{u'v'}/\Delta U^2$ outside the nominal mixing-layer edge, particularly near the splitter plate in the near field. 
These negative excursions likely originate from the localized wake generated by the vane end plates and its strong influence close to the splitter plate trailing edge (see \S~\ref{sec:MeanVelocity}, figures~\ref{fig:Figure5} and \ref{fig:Figure6}). 
Their confinement to the near-field region further supports this interpretation.

The planar and skewed mixing layers exhibit qualitatively similar shear-stress distributions overall. 
The planar configuration, however, remains approximately symmetric about the mixing-layer centerline, except near the splitter plate at $x/h=0.5$. 
In contrast, the skewed configuration exhibits pronounced asymmetry in the near field, with the peak Reynolds shear stress displaced toward the low-speed side of the flow. 
Farther downstream, the profiles progressively recover approximate symmetry.

This behavior closely mirrors the asymmetry observed previously in the mean-vorticity distributions discussed in \S~\ref{sec:velGrad} and shown in figure~\ref{fig:Figure8}. 
The physical origin of this asymmetry is not immediately obvious. 
Nevertheless, several earlier investigations of planar mixing layers~\citep{bell1990development,azim2003plane} reported similar low-speed-side biasing and attributed it to splitter-plate wake effects localized on the low-speed side (see \S~\ref{sec:MeanVelocity}). 
The present observations appear to support that interpretation. 
In particular, the stronger and more persistent asymmetry observed in the skewed configuration suggests that the imposed mean-flow three-dimensionality amplifies the influence of the splitter-plate wake during the transient development of the mixing layer. 

The Reynolds shear-stress profiles for both configurations progressively approach a Gaussian-like form with increasing downstream distance, consistent with previous experimental and numerical observations~\citep{loucks2012velocity,azim2003plane,bell1990development}. 
This behavior can be interpreted within the framework of an eddy-viscosity approximation. 
Using the Boussinesq hypothesis,
$-\overline{u'v'}=
\nu_t{\partial U}/{\partial y}$,
together with Prandtl's constant-eddy-viscosity model,
$\nu_t=c_1\delta_U\Delta U$, one obtains~\citep{boussinesq1877essai,prandtl1942bemerkungen}

\begin{equation}
\frac{\overline{u'v'}}{\Delta U^2} =
\frac{c_1}{\sigma\sqrt{2\pi}}\exp\left[-
\left(\frac{\eta}{\sigma\sqrt{2}}
\right)^2\right]
\label{eqn:Restress}
\end{equation}
where we used the self-similar mean-velocity gradient given by \eqref{eqn:velGrad}. 

For an error-function representation of the mean velocity profile, the Reynolds shear-stress distribution attains its maximum at the mixing-layer centerline ($\eta=0$). 
Under the constant eddy-viscosity approximation, the peak normalized Reynolds shear stress becomes~\citep{Pope_2000}

\begin{equation}
\frac{\overline{u'v'}_{\max}}{\Delta U^2}
=
\frac{d\delta/dx}{\lambda}
\frac{\sigma}{2\sqrt{2\pi}}
\label{eqn:maxRestressprofile}
\end{equation}
The corresponding eddy-viscosity coefficient is therefore

\begin{equation}
c_1=\frac{d\delta/dx}{\lambda}\frac{\sigma^2}{2}.
\label{eqn:Coeff_eddyVisc}
\end{equation}

Figure~\ref{fig:Figure9}(\textit{a,b}) compares the analytical form given above with the measured Reynolds shear-stress profiles, with the companion LES, and with previously reported data from experiments~\citep{bell1990development} and direct numerical simulations~\citep{rogers1994direct}. 
The model reproduces the approximately Gaussian profile shape observed in both the present measurements and the existing literature reasonably well. 
The predicted peak amplitude from \eqref{eqn:maxRestressprofile}, however, exceeds the canonical asymptotic self-similar value by approximately $30\%$. 
Consequently, the analytical distribution remains broader and encompasses the full range of experimentally observed profiles, including those obtained in the present transient regime. 
If, instead, we determine the model coefficient using the empirically observed asymptotic peak value, $|\overline{u'v'}_{\max}|/\Delta U^2 \approx 0.011$,
reported in previous experimental and numerical studies~\citep{yoder2015modeling,bell1990development,rogers1994direct}, the resulting analytical distribution reproduces the width of the expected self-similar Reynolds shear-stress profile remarkably well. 

To compare the maximum stress magnitudes quantitatively, we examined the downstream evolution of the peak Reynolds shear stress, $\overline{u'v'}_{\max}/\Delta U^2$, shown in figure~\ref{fig:Figure9}(\textit{c}). 
For both configurations, the peak stress increases progressively with downstream distance throughout the region of interest. 
However, the skewed mixing layer consistently exhibits smaller peak values than the planar case, with reductions of up to approximately 65\% (figure~\ref{fig:Figure9}(\textit{d})). The companion LES exhibits a similar reduction (figure~\ref{fig:Figure9}(\textit{a,b}), suggesting that this behavior is a robust feature of the skewed configuration rather than a consequence of the specific experimental arrangement.
Neither case attains the canonical asymptotic value of approximately $0.011$ typically reported for fully developed planar turbulent mixing layers \citep{yoder2015modeling,bell1990development,rogers1994direct}. 

\subsubsection{Secondary shear stress}\label{sec:secondaryShear}

\begin{figure}
 \centering

 \includegraphics[width=1\linewidth,trim={35 30 40 60},clip]{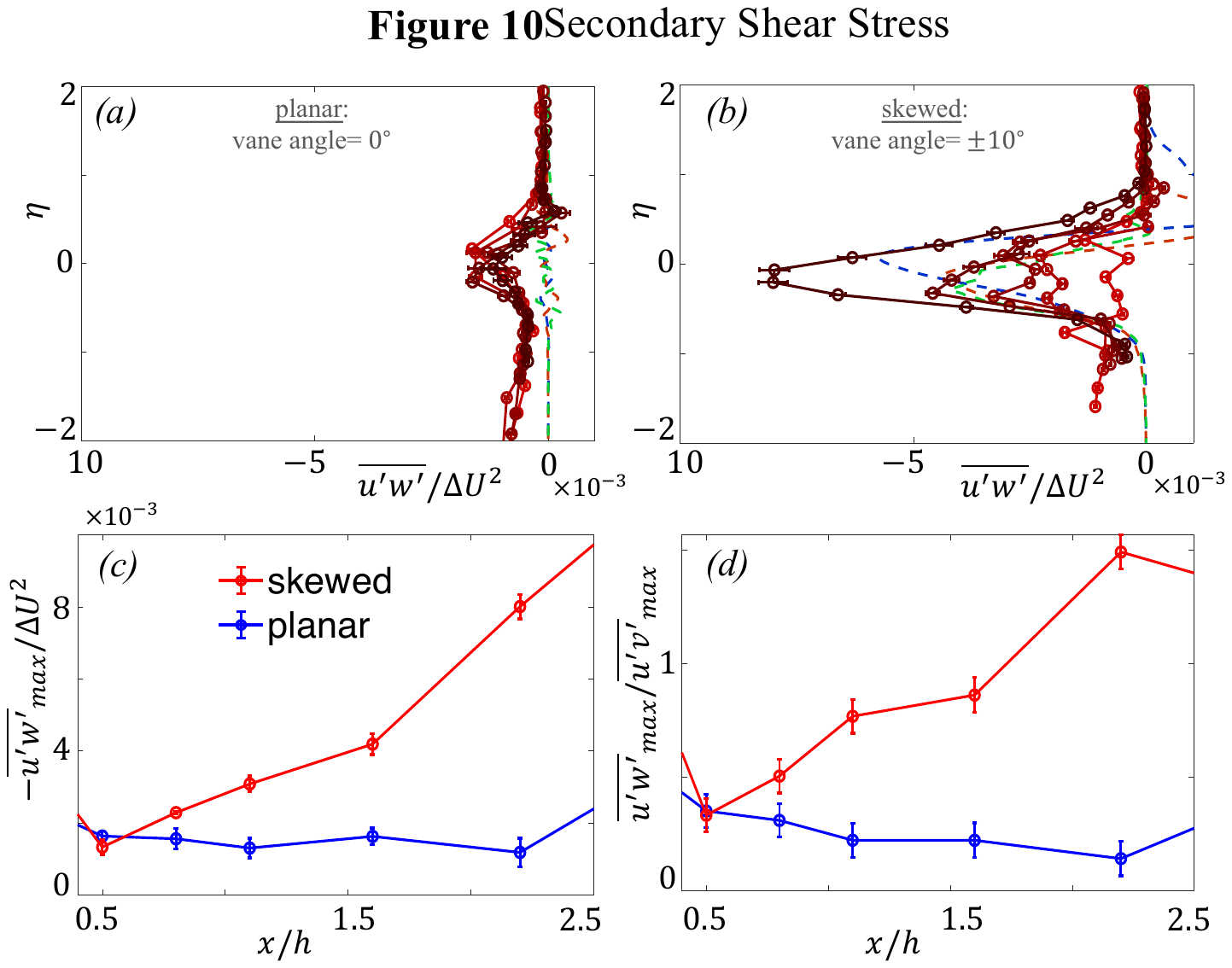}
\caption{Profiles of secondary Reynolds shear stress. 
\textit{(a)} Planar mixing layers. 
\textit{(b)} Skewed mixing layers. 
Legends are as in figure~\ref{fig:Figure9}. 
\textit{(c)} Streamwise evolution of the normalized maximum secondary shear stress; blue and red markers correspond to planar and skewed cases, respectively. 
\textit{(d)} Downstream evolution of the ratio of the normalized maximum secondary shear stress to the normalized primary shear stress.}
 \label{fig:Figure10}
\end{figure}

Figure~\ref{fig:Figure10}(\textit{a,b}) shows the profiles of the secondary Reynolds shear stress, $\overline{u'w'}/\Delta U^2$,
at different downstream locations for both the planar and skewed mixing layers within the region of interest. 
In the skewed configuration, the imposed directional shear naturally generates correlations between the streamwise and spanwise velocity fluctuations, and therefore a non-zero secondary shear stress is expected. 
In contrast, for an ideal planar turbulent mixing layer, statistical symmetry implies that $\overline{u'w'}$ should vanish identically \citep{Pope_2000}. 

The present measurements show that the planar configuration exhibits nonzero secondary shear stress within the central mixing region. 
The profiles display a distinct bell-shaped distribution, with peak magnitude near the center of the mixing layer. 
Similar observations were previously reported in \citet{bell1989three} and \citet{bell1992measurements}, which associate the non-zero $\overline{u'w'}$ in planar mixing layers with the presence of streamwise vorticity generated by three-dimensional braid instabilities. 
For both planar and skewed configurations, the secondary stress decays approximately to zero on the low-speed side of the mixing layer. 
On the high-speed side, however, $\overline{u'w'}$ remains nonzero, although its magnitude becomes comparatively small. 

For the skewed mixing layer, this asymmetry may partly reflect the large-scale circulatory motion associated with the return flow along the high-speed side of the tunnel cross-section, as discussed in \S~\ref{sec:RoleofTurningVanes}. 
The persistence of nonzero $\overline{u'w'}$ on the high-speed side even in the planar configuration, however, indicates that circulation alone cannot account for this behavior. 
One possible contributing factor is the asymmetry in the inflow turbulence conditions introduced by the fine mesh screen used to generate the low-speed stream. 
The mesh likely suppresses turbulence intensity on the low-speed side, whereas the absence of a screen on the high-speed side permits larger incoming turbulent fluctuations. 
Nevertheless, the precise mechanism responsible for the persistence of the secondary shear stress outside the nominal mixing region remains unclear at the time of writing.

Figure~\ref{fig:Figure10}(\textit{c}) shows the downstream evolution of the peak secondary Reynolds shear stress. 
For the planar mixing layer,
$\overline{u'w'}_{\max}/\Delta U^2$ remains approximately constant throughout the region of interest, with a magnitude of approximately $ 2\times10^{-3}$. 
In contrast, the skewed mixing layer exhibits a monotonic downstream increase in the peak secondary stress. The companion LES also yields nonzero values of the secondary Reynolds shear stress in the skewed configuration, indicating that this behavior, like the reduction in the primary Reynolds shear stress, is a robust feature of the skewed flow rather than a consequence of the particular experimental arrangement.
To compare the relative importance of the primary and secondary shear stresses, figure~\ref{fig:Figure10}(\textit{d}) shows the ratio
$\overline{u'w'}_{\max}/\overline{u'v'}_{\max}$.
At the nearest downstream station ($x/h=0.5$), both configurations exhibit
$\overline{u'w'}_{\max}/\overline{u'v'}_{\max}
\approx0.4$,
which agrees reasonably well with the value of approximately $0.5$ reported in \citet{bell1989three} for planar mixing layers. 
This work further reports a ratio that decreases downstream and becomes negligible once the flow approaches a self-similar state in which braid-induced streamwise vorticity weakens substantially. 
The present planar mixing-layer measurements exhibit the same trend, with the ratio decreasing to approximately $0.2$ by $x/h=2.2$. 
These observations support the interpretation that the nonzero secondary shear stress in planar mixing layers originates from streamwise vorticity associated with transient three-dimensional braid dynamics.

The skewed mixing layer, however, exhibits different behavior. 
Instead of decaying downstream, the ratio
$\overline{u'w'}_{\max}/\overline{u'v'}_{\max}$
increases monotonically from approximately $0.4$ near the splitter plate to values approaching $1.5$ at the furthest downstream location. 
This downstream amplification cannot plausibly result from braid-induced streamwise vorticity alone, since the streamwise vortices generated by secondary instabilities are expected to weaken rather than intensify downstream. 
Instead, the present observations strongly suggest the existence of an additional source of streamwise vorticity associated directly with the imposed directional shear. 
We therefore attribute the sustained growth of the secondary shear stress in the skewed mixing layer to the persistent streamwise-oriented mean vorticity generated by the imposed spanwise velocity gradients discussed previously in \S~\ref{sec:velGrad}.

\subsubsection{Normal stresses and turbulent kinetic energy}\label{sec:3.2.3}

\begin{figure}
 \centering
 \includegraphics[width=1\linewidth,trim={20 30 20 70},clip]{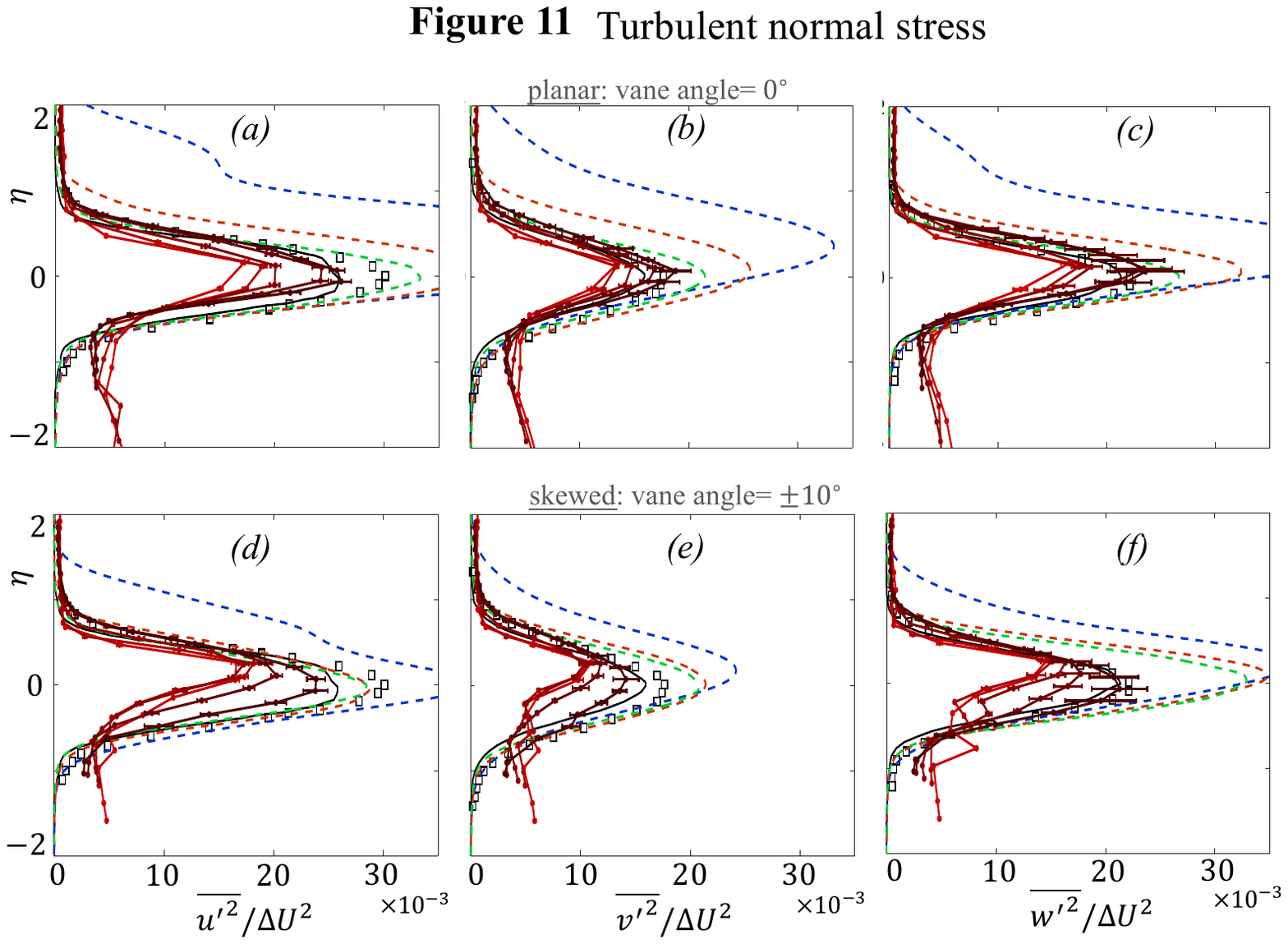}
\caption{Profiles of Normal stresses. 
The top panels \textit{(a--c)} are for planar mixing layers while the bottom panels \textit{(d--f)} are for skewed cases. 
\textit{(a,d)} $\overline{u'^2}/\Delta U^2$, \textit{(b,e)} $\overline{v'^2}/\Delta U^2$, and \textit{(c,f)} $\overline{w'^2}/\Delta U^2$. 
Legends as in Figure~\ref{fig:Figure9}.}
 \label{fig:Figure11}

\end{figure}

Figure~\ref{fig:Figure11}(\textit{a--f}) shows the profiles of the normal Reynolds stresses, namely the streamwise component $\overline{u'^2}$, the cross-stream component $\overline{v'^2}$, and the spanwise component $\overline{w'^2}$, for both planar and skewed mixing layers within the region of interest. 
For both configurations, the normalized normal-stress profiles exhibit approximately Gaussian-like shapes across the mixing region. 
The mixing-layer profiles compare reasonably well with previously reported measurements and DNS results for self-preserving planar mixing layers \citep{bell1990development,rogers1994direct}. Agreement with the companion LES is more limited and is observed primarily at the later stages of evolution. On the high-speed side, by contrast, nonzero residual stress levels persist beyond the nominal edge of the layer, with average magnitudes of approximately $0.005~\Delta U^2$. 
A qualitatively similar behavior was also observed for the secondary shear stress in \S~\ref{sec:secondaryShear} and as discussed previously, this asymmetry may partly originate from the specific inflow conditions used to generate the two freestreams.

For the skewed configuration, the normal-stress profiles initially exhibit a clear asymmetry, with their peak values displaced toward the low-speed side, before progressively recovering a more symmetric distribution farther downstream. Similar asymmetric profiles are also observed in the companion LES during the early stages of evolution for both the planar and skewed cases. 
In addition, all three normal stresses in the present study display a distinct secondary peak near $\eta \approx -0.4$, corresponding approximately to the vane-height location. 
This secondary peak decays rapidly beyond $x/h \approx 0.8$ and is either weak or absent in the planar case. 
Because this feature appears in the same region as the strongest end-plate wake effects (see figure~\ref{fig:Figure5} and \S~\ref{sec:MeanVelocity}), the observations suggest that the imposed mean flow three-dimensionality amplifies the wake-induced perturbations generated by the vane-end-plate assembly. 

\begin{figure}
 \centering
 \includegraphics[width=1\linewidth,trim={15 210 20 98},clip]{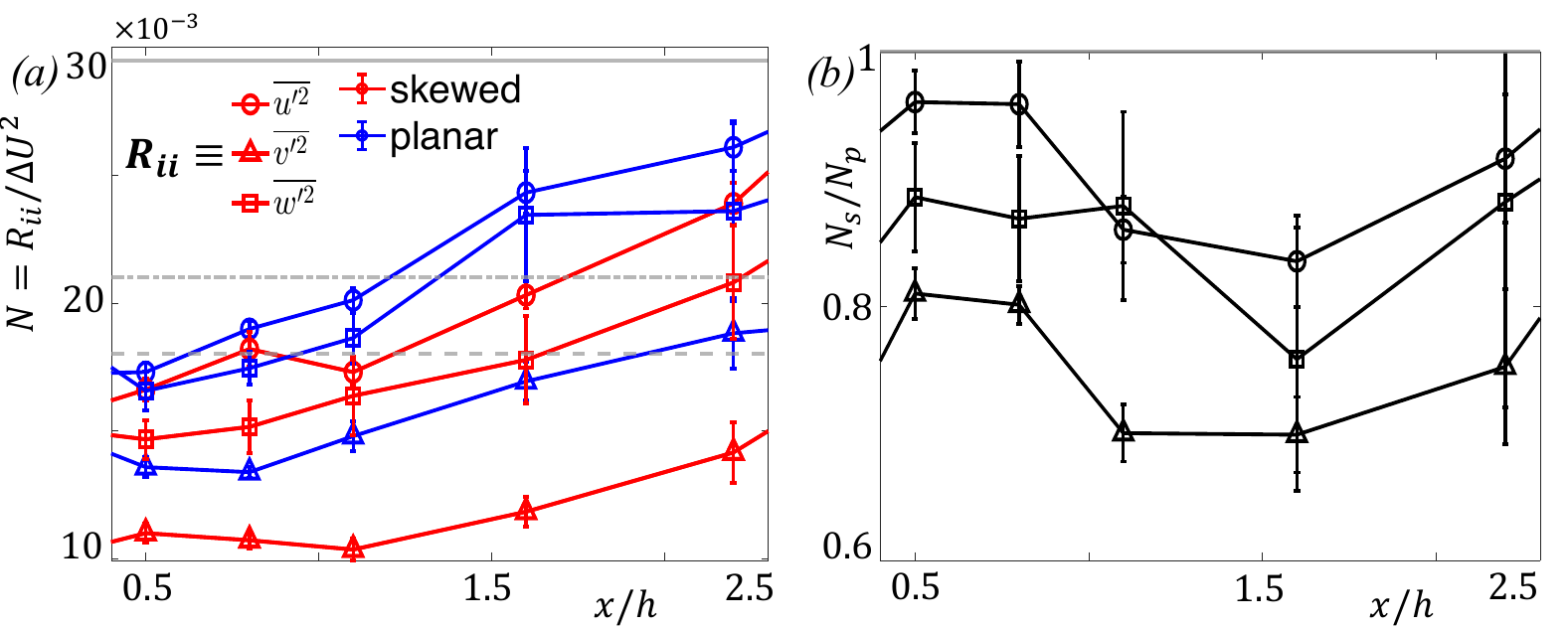}
\caption{ \textit{(a)} Streamwise evolution of the normalized maximum normal stresses, $N$. 
Blue and red markers correspond to planar and skewed mixing layers, respectively. 
The grey curves denote the maximum values reported in the literature for self-similar planar mixing layers \citep{yoder2015modeling}: solid, $\overline{u'^2}{\max}/\Delta U^2$; dashed, $\overline{v'^2}{\max}/\Delta U^2$; and dash-dotted, $\overline{w'^2}_{\max}/\Delta U^2$. 
\textit{(b)} Streamwise evolution of the ratio of the normalized maximum normal stresses in the skewed configuration ($N_s$) to those in the planar configuration ($N_p$).}
 \label{fig:Figure12}
\end{figure}

Figure~\ref{fig:Figure12}(\textit{a}) shows the streamwise evolution of the normalized peak normal stresses for both planar and skewed mixing layers. 
In all cases, the peak stress magnitudes increase with downstream distance over the investigated region of interest and do not exhibit a tendency toward saturation. 
The only partial exception is the planar $\overline{w'^2}_{\max}/\Delta U^2$, which remains approximately constant beyond $x/h \approx 1.5$. 
For all measured downstream stations within the region of interest, the skewed mixing layer exhibits smaller values of peak normal stresses than the planar case, with reductions by approximately 30\% (see figure~\ref{fig:Figure12}(\textit{b}). The companion LES displays a qualitatively similar observation (figure~\ref{fig:Figure11}), providing independent support for this observation.

For a quantitative comparison of the normalized peak normal stresses, figure~\ref{fig:Figure12}(\textit{a}) includes reference values representative of asymptotic self-preserving state of fully developed planar mixing layers, compiled in \citet{yoder2015modeling} from existing experimental and numerical studies. 
Consistent with canonical mixing-layers, the streamwise normal stress remains the dominant component, followed by the spanwise and cross-stream components. 
Relative to these self-preserving reference values, all peak normal stresses in the skewed case remain systematically lower throughout the investigated region. 
In the planar case, however, only $\overline{u'^2}_{\max}/\Delta U^2$ remains below the expected asymptotic level, whereas $\overline{v'^2}_{\max}/\Delta U^2$ and $\overline{w'^2}_{\max}/\Delta U^2$ exceed the nominal self-similar range. 
Among the three components, only $\overline{w'^2}_{\max}/\Delta U^2$ approaches an approximately constant value near $0.023$, consistent with reported self-similar values of $0.021$, whereas $\overline{v'^2}_{\max}/\Delta U^2$ continues to increase beyond the expected asymptotic range of $0.018$.

\begin{figure}
 \centering

 \includegraphics[width=1\linewidth,trim={20 50 40 60},clip]{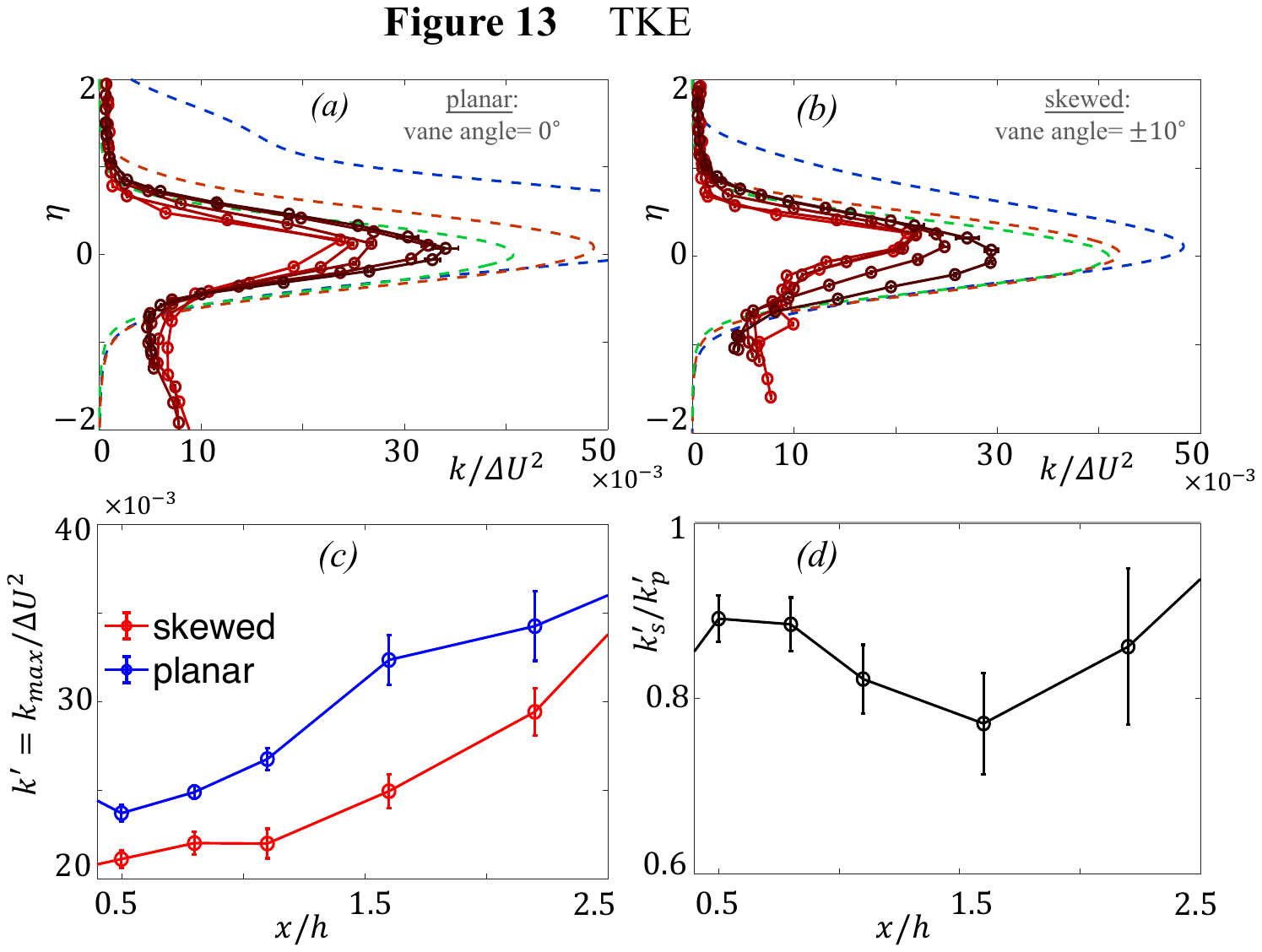}
\caption{ Profiles of total kinetic energy. 
\textit{(a)} Planar mixing layers. 
\textit{(b)} Skewed mixing layers. 
Legends as in Fig.~\ref{fig:Figure8}. 
\textit{(c)} Streamwise development of the normalized maximum values, $k'$. 
\textit{(d)} Streamwise evolution of the ratio of the normalized maxima in the skewed configuration ($k'_s$) to those in the planar configuration ($k'_p$). }
 \label{fig:Figure13}
\end{figure}

Figure~\ref{fig:Figure13}(\textit{a,b}) shows the profiles of the turbulent kinetic energy, $k=(\overline{u'^2}+\overline{v'^2}+\overline{w'^2})/2$,
for both planar and skewed mixing layers. 
Outside the mixing region, $k$ decays to nearly zero on the low-speed side, whereas nonzero residual values persist on the high-speed side, consistent with the behavior previously observed for the normal stresses and the secondary Reynolds shear stress.

The skewed mixing layer exhibits an initially asymmetric distribution, with the peak displaced toward the low-speed side, before progressively recovering a more symmetric profile further downstream. 
Compared with the planar case, the skewed configuration also displays a somewhat narrower spatial distribution of turbulent kinetic energy within the central mixing region.

The maximum turbulent kinetic energy increases with downstream distance throughout the investigated region for both configurations. 
However, the planar mixing layer consistently attains larger peak values than the skewed case, with differences reaching approximately $20\%$ (figure~\ref{fig:Figure13}(\textit{c,d})). A qualitatively similar observation is found in the companion LES (figure~\ref{fig:Figure13}(\textit{a,b})), where the skewed configuration likewise exhibits reduced peak turbulent kinetic energy relative to the planar case.

According to the commonly accepted criteria for asymptotic evolution in turbulent mixing layers \citep{townsend1976structure}, namely, collapse of the normalized mean-velocity profiles, linear downstream growth of the characteristic thickness scales, collapse of the normalized Reynolds-stress distributions, and invariance of their peak amplitudes with increasing downstream distance, the present results indicate that both the planar and skewed configurations attain an approximately self-preserving mean-flow state within the investigated region of interest, but do not reach a self-preserving turbulent state with respect to the normal and shear stresses statistics. 
In particular, although the mean velocity profiles and characteristic thicknesses exhibit behavior broadly consistent with asymptotic evolution, the normal and shear stresses and turbulent kinetic energy continue to evolve significantly throughout the region of interest. 
Such a disparity between the development of the mean flow and that of the turbulence field is well known in turbulent shear flows and has previously been reported for planar mixing layers \citep{bell1990development,azim2003plane}.

\citet{bradshaw1966effect} estimates that the downstream distance required to reach a fully self-preserving turbulent state is approximately $1000~\theta_0$. 
Although the analysis presented therein focused on a single-stream mixing layer, subsequent studies, including \citet{brown1974density}, suggested that a comparable distance applies also to two-stream mixing layers. 
Using the sum of the initial momentum thicknesses, $(\theta_{0H}+\theta_{0L})$, for the present configuration yields an estimated self-preservation distance of approximately $x/h \approx 15$. 
Similarly, the criterion proposed in \citet{ho1984perturbed}, \textit{i.e.}, $\lambda x/\lambda_n \approx 10$ with $\lambda_n=\theta_0/0.032$, gives an estimated development length of approximately $x/h \approx 11.5$. 
Both estimates substantially exceed the investigated downstream range, $0.5 \leq x/h \lesssim 2$, demonstrating that the present measurements probe an extended transient regime rather than an asymptotic self-preserving state.

\subsubsection{Efficiency of turbulent transport: Townsend's structure parameter}\label{sec:TownsendStr}

\begin{figure}
 \centering

 \includegraphics[width=1\linewidth,trim={20 30 40 60},clip]{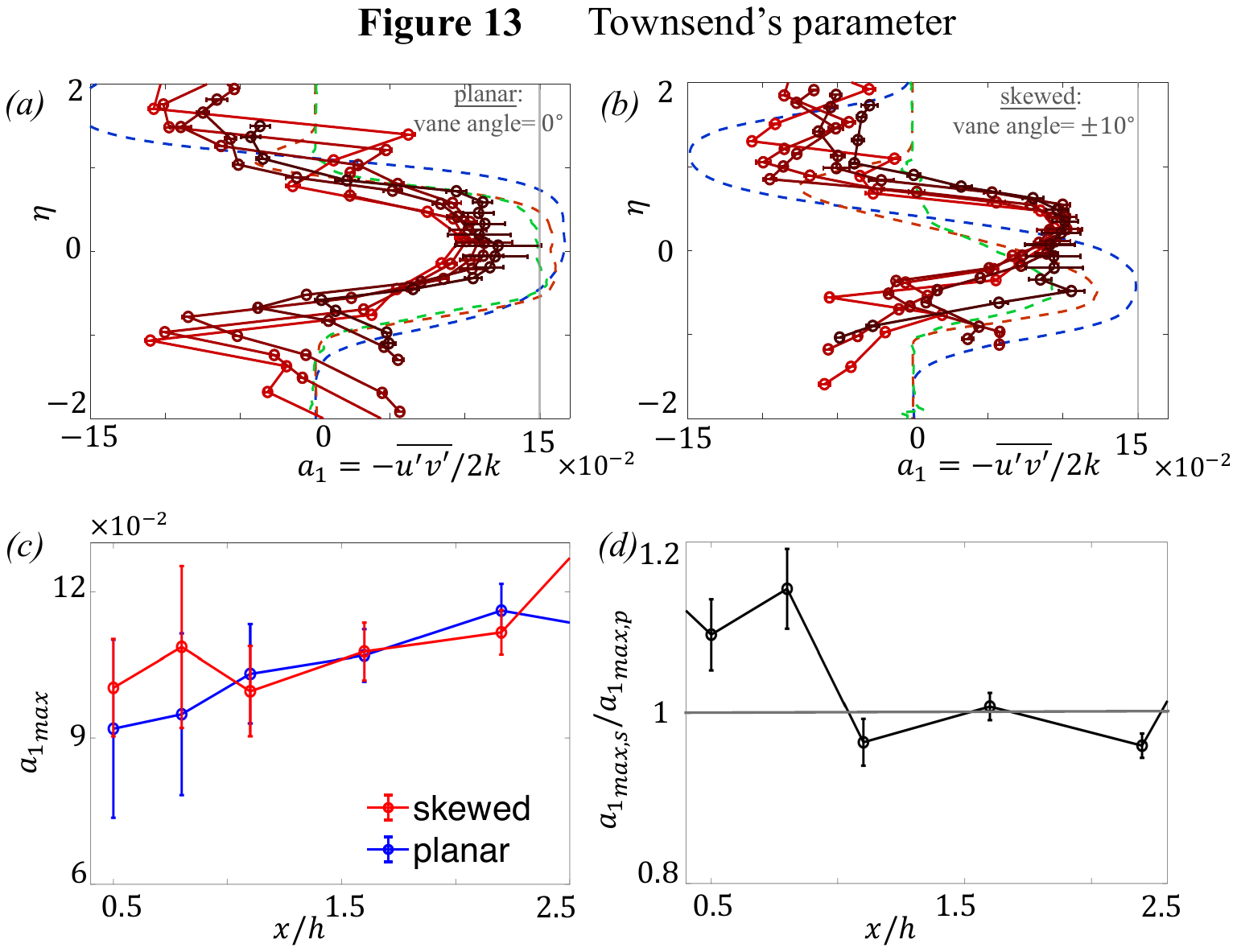}
\caption{ Townsend's structure parameter, $a_1$. 
\textit{(a)} Planar mixing layers. 
\textit{(b)} Skewed mixing layers. 
Legends as in Fig. \ref{fig:Figure9}. 
\textit{(c)} Streamwise development of the normalized maximum values, $a_{1\max}$. 
\textit{(d)} Streamwise evolution of the ratio of the normalized maximum in the skewed configuration ($a_{1\max,\mathrm{s}}$) to those in the planar configuration ($a_{1\max,\mathrm{s}}$).}

 \label{fig:Figure14}
\end{figure}

One might expect that the imposed mean-flow three-dimensionality would enhance the efficiency of lateral momentum transport relative to the canonical planar mixing layer. 
In studies of three-dimensional turbulent boundary layers (3DTBLs), this efficiency is commonly quantified using Townsend’s structure parameter, $a_1$, defined as $a_1 = (\overline{u'v'} + \overline{v'w'})/(2k)$, which relates the Reynolds shear stress to the turbulent kinetic energy and therefore measures the effectiveness with which turbulent fluctuations sustain momentum transport \citep{townsend1976structure,johnston1996advances}. 
Figure~\ref{fig:Figure14}(\textit{a--b}) presents the profiles of $a_1$ for both planar and skewed mixing layers within the region of interest. 
For a broad class of turbulent shear flows, including canonical planar mixing layers, $a_1$ typically approaches values near $0.15$ in the fully self-preserving state \citep{townsend1976structure}. 

For both planar and skewed mixing layers, the profiles of $a_1$ collapse reasonably well within the central mixing region to approximately constant values. 
However, the measured values remain systematically below the canonical asymptotic value of $a_1 \approx 0.15$ for both configurations. 
Owing to the limitations of the hot-wire arrangement, the cross-shear term $\overline{v'w'}$ could not be measured directly and was therefore omitted from the experimental estimate, reducing the evaluated parameter to the effective form $a_1 \approx \overline{u'v'}/(2k)$. 
For the planar configuration, however, $\overline{v'w'}$ is expected to vanish by symmetry. 
Moreover, the companion LES indicates that the contribution of $\overline{v'w'}$ remains below approximately $2\%$ even in the skewed configuration. 
The observed discrepancy relative to the asymptotic reference value therefore cannot be explained by omission of the cross-shear term. Interestingly, the companion LES approaches the expected asymptotic value, at least for the planar case, at long times.
This suggests that the comparatively lower values measured experimentally are more likely a consequence of the flow remaining in a transient regime rather than having reached a fully self-preserving turbulent state. 
Such an interpretation is consistent with the continued downstream evolution of several turbulence statistics discussed in the preceding sections.

The planar configuration yields $a_{1\max}\approx0.11\pm0.02$, whereas the skewed configuration gives $a_{1\max}\approx0.10\pm0.02$ (figure~\ref{fig:Figure14}(\textit{c})). 
Their ratio remains close to unity throughout the region of interest (figure~\ref{fig:Figure14}(\textit{d})). 
These results suggest that, within experimental uncertainty, the imposed mean-flow three-dimensionality does not significantly alter the efficiency with which turbulent kinetic energy sustains the primary Reynolds shear stress. 
This behavior contrasts with observations in 3DTBLs, where $a_1$ decreases systematically with increasing mean-flow skew angle \citep{johnston1996advances,lozano2020non}. 
For skew angles comparable to the effective $20^\circ$ imposed in the present study, previous investigations of 3DTBLs reported reductions in $a_1$ of approximately $30\%$ relative to the corresponding two-dimensional boundary layer \citep{johnston1996advances}.

\section{Overall Summary and Conclusion}\label{sec:conclusion}

We developed and demonstrated a simple and easily realizable framework to generate controlled mean-flow three-dimensionality in turbulent mixing layers using turning vanes mounted near the trailing edge of a splitter plate. 
We then employed this framework to perform a comparative experimental investigation of planar and skewed turbulent mixing layers in a wind-tunnel facility and to examine how imposed mean-flow skew modifies the evolution of free-shear turbulence relative to the canonical planar case. 
The experiments generated two parallel streams of unequal velocity using a passive turbulence-generating grid partially covered with fine mesh.  
A fixed vane angle of $\pm 10^\circ$ produced the skewed configuration, while alignment with the freestream recovered the planar case. 
The results demonstrate that the approach provides a practical framework to systematically study the influence of imposed mean-flow three-dimensionality on turbulent free-shear flows.

Quantitatively, the skewed configuration exhibits systematically lower magnitudes of mean velocity, velocity gradients, and primary shear and normal stresses relative to the planar case. 
In contrast, the mixing-layer thickness and the secondary shear stress increase in the skewed flow. 

Despite these differences, several key normalized quantities remain qualitatively unaltered. Both configurations display similar streamwise evolution: the mean velocity profiles follow an error-function form, the layer thickness grows linearly at comparable rates, and the Reynolds-stress distributions retain approximately Gaussian shapes with maxima located within the mixing region. 
Most notably, Townsend’s structure parameter, $a_1$, which characterizes the efficiency of turbulent momentum transport relative to the turbulent kinetic energy, remains invariant within statistical uncertainty between the planar and skewed configurations. 
This behavior contrasts with 3D TBLs, where comparable levels of mean-flow skewing produce reductions of $\approx 30\%$ in $a_1$. 

The origin of the observed quantitative differences, however, remains unresolved. 
An important open question is whether these differences arise primarily from the reduced mean shear or from the imposed spanwise velocity component itself. 
Previous studies provide partial insight. 
\citet{mehta1991effect} shows that variations in velocity ratios modify the magnitude of mean and turbulent quantities in planar mixing layers without significantly altering their qualitative structure or asymptotic behaviour, while \citet{meldi2020numerical} demonstrates using direct numerical simulations that mean skew predominantly influences the near field, whereas velocity shear governs the fully developed state. 
However, previous investigations have largely focused either on canonical planar mixing layers or on skewed mixing layers spanning only a limited range of skew angles and initialized using idealized hyperbolic-tangent mean profiles. 
These limitations motivate future studies over broader parameter ranges, including skew angles and velocity ratios, to determine whether sufficiently strong mean-flow three-dimensionality can fundamentally reorganize mixing-layer dynamics.

Further insight into the observed quantitative differences may be obtained by examining the evolution of the flow structures. 
Flow-visualization techniques such as shadowgraphy, particle image velocimetry (PIV), or Schlieren imaging could be used to characterize the large-scale coherent motions and to determine how imposed mean-flow skewness modifies their organization. 
In particular, such measurements could clarify whether the skewed mixing layer, like the planar configuration, supports spanwise rollers and streamwise vortical structures, and whether these coherent motions retain similar dynamical characteristics in both cases. 
In addition to the large-scale structures, a comparative investigation of the fine-scale turbulence dynamics in the planar and skewed configurations is currently underway and may provide further insight into the persistence of anisotropy across scales. 
Such measurements could help determine whether the large-scale quantitative differences observed here extend down to the smallest dissipative scales of the turbulence.

In summary, the present study demonstrates that controlled mean-flow three-dimensionality can be generated in turbulent mixing layers using turning-vanes while maintaining experimental simplicity. 
Although the imposed skew produces measurable changes in both mean and turbulent statistics, the dominant structural and transport characteristics of the mixing layer appear to remain largely unchanged. 
Beyond these physical findings, the study establishes an experimental framework and provides an empirical benchmark for future investigations of three-dimensional free-shear turbulence. 


\begin{bmhead}[Acknowledgement.]
DG and GPB are grateful to Profs.~Edwin (Todd) Cowen, Olivier Desjardins, Tobias Schneider and Zellman Warhaft for discussions and suggestions, as well as to Grace Fischler, Danielle Fulep, Fariz Kabir, Peter McGurk, Anthony Ordonez, and Runxuan Li for their assistance with the experimental setup during the initial stages of the project. 

\end{bmhead}

\begin{bmhead}[Funding.]
The work was supported through
Office of Naval Research Grant (N00014-22-1-2038)
\end{bmhead}

\begin{bmhead}[Declaration of Interests.]
The authors report no conflict of interest.
\end{bmhead}

\bibliographystyle{jfm}
\bibliography{jfm_bibliography}%

\begin{thebibliography}{66}
\expandafter\ifx\csname natexlab\endcsname\relax\def\natexlab#1{#1}\fi
\def\au#1{#1} \def\ed#1{#1} \def\yr#1{#1}\def\at#1{#1}\def\jt#1{\textit{#1}} \def\bt#1{#1}\def\bvol#1{\textbf{#1}} \def\vol#1{#1} \def\pg#1{#1} \def\publ#1{#1}\def\arxiv#1{#1}\def\org#1{#1}\def\st#1{\textit{#1}}

\bibitem[Abbott \& Von~Doenhoff(2012)]{abbott2012theory}
{\sc \au{Abbott, Ira~H} \& \au{Von~Doenhoff, Albert~E}} \yr{2012} {\em Theory of wing sections: including a summary of airfoil data\/}.  \publ{Courier Corporation}.

\bibitem[Abramovi{\v{c}}(1963)]{abramovivc1963theory}
{\sc \au{Abramovi{\v{c}}, Genrih~Naumovi{\v{c}}}} \yr{1963} {\em The theory of turbulent jets\/}.  \publ{MIT Press}.

\bibitem[Azim \& Islam(2003)]{azim2003plane}
{\sc \au{Azim, MA} \& \au{Islam, AKMS}} \yr{2003}  \at{Plane mixing layers from parallel and non-parallel merging of two streams}.  \jt{Experiments in Fluids}  \bvol{34}~(2),  \pg{220--226}.

\bibitem[Bell \& Mehta(1989)]{bell1989three}
{\sc \au{Bell, James} \& \au{Mehta, Rabindra}} \yr{1989} Three-dimensional structure of a plane mixing layer.  \bt{In {\em 27th Aerospace Sciences Meeting\/}},  \pg{p. 124}.

\bibitem[Bell \& Mehta(1990)]{bell1990development}
{\sc \au{Bell, James~H} \& \au{Mehta, Rabindra~D}} \yr{1990}  \at{Development of a two-stream mixing layer from tripped and untripped boundary layers}.  \jt{AIAA journal}  \bvol{28}~(12),  \pg{2034--2042}.

\bibitem[Bell \& Mehta(1992)]{bell1992measurements}
{\sc \au{Bell, James~H} \& \au{Mehta, Rabindra~D}} \yr{1992}  \at{Measurements of the streamwise vortical structures in a plane mixing layer}.  \jt{Journal of Fluid Mechanics}  \bvol{239},  \pg{213--248}.

\bibitem[Boldman {\em et~al.\/}(1976)Boldman, Brinich \& Goldstein]{boldman1976vortex}
{\sc \au{Boldman, DR}, \au{Brinich, PF} \& \au{Goldstein, ME}} \yr{1976}  \at{Vortex shedding from a blunt trailing edge with equal and unequal external mean velocities}.  \jt{Journal of Fluid Mechanics}  \bvol{75}~(4),  \pg{721--735}.

\bibitem[Boukharfane {\em et~al.\/}(2021)Boukharfane, Er-raiy, Elkarii \& Parsani]{boukharfane2021direct}
{\sc \au{Boukharfane, Radouan}, \au{Er-raiy, Aimad}, \au{Elkarii, M} \& \au{Parsani, Matteo}} \yr{2021}  \at{A direct numerical simulation study of skewed three-dimensional spatially evolving compressible mixing layer}.  \jt{Physics of Fluids}  \bvol{33}~(11).

\bibitem[Boussinesq(1877)]{boussinesq1877essai}
{\sc \au{Boussinesq, Joseph}} \yr{1877} {\em Essai sur la th{\'e}orie des eaux courantes\/}.  \publ{Impr. nationale}.

\bibitem[Bradshaw(1966)]{bradshaw1966effect}
{\sc \au{Bradshaw, P}} \yr{1966}  \at{The effect of initial conditions on the development of a free shear layer}.  \jt{Journal of Fluid Mechanics}  \bvol{26}~(2),  \pg{225--236}.

\bibitem[Bradshaw \& Pontikos(1985)]{bradshaw1985measurements}
{\sc \au{Bradshaw, P} \& \au{Pontikos, NS}} \yr{1985}  \at{Measurements in the turbulent boundary layer on an ‘infinite’swept wing}.  \jt{Journal of Fluid Mechanics}  \bvol{159},  \pg{105--130}.

\bibitem[Braud {\em et~al.\/}(2004)Braud, Heitz, Arroyo, Perret, Delville \& Bonnet]{braud2004low}
{\sc \au{Braud, Caroline}, \au{Heitz, Dominique}, \au{Arroyo, Georges}, \au{Perret, Laurent}, \au{Delville, Jo{\"e}l} \& \au{Bonnet, Jean-Paul}} \yr{2004}  \at{Low-dimensional analysis, using pod, for two mixing layer--wake interactions}.  \jt{International Journal of Heat and Fluid Flow}  \bvol{25}~(3),  \pg{351--363}.

\bibitem[Browand \& Troutt(1985)]{browand1985turbulent}
{\sc \au{Browand, FK} \& \au{Troutt, TR}} \yr{1985}  \at{The turbulent mixing layer: geometry of large vortices}.  \jt{Journal of Fluid Mechanics}  \bvol{158},  \pg{489--509}.

\bibitem[Brown \& Roshko(1974)]{brown1974density}
{\sc \au{Brown, GL} \& \au{Roshko, A}} \yr{1974}  \at{On density effects and large structure in turbulent mixing layers}.  \jt{Journal of Fluid Mechanics}  \bvol{64}~(4),  \pg{775--816}.

\bibitem[Bruun(1995)]{Bruun1995}
{\sc \au{Bruun, H~H}} \yr{1995} {\em Hot-Wire Anemometry: Principles and Signal Analysis\/}.  \publ{Oxford University Press}.

\bibitem[Chandrsuda {\em et~al.\/}(1978)Chandrsuda, Mehta, Weir \& Bradshaw]{chandrsuda1978effect}
{\sc \au{Chandrsuda, C}, \au{Mehta, Rabindra~D}, \au{Weir, AD} \& \au{Bradshaw, Peter}} \yr{1978}  \at{Effect of free-stream turbulence on large structure in turbulent mixing layers}.  \jt{Journal of Fluid Mechanics}  \bvol{85}~(4),  \pg{693--704}.

\bibitem[Dimotakis(1989)]{dimotakis1989turbulent}
{\sc \au{Dimotakis, Paul~E}} \yr{1989} Turbulent free shear layer mixing and combustion.  \bt{In {\em 9th International Symposium on Air Breathing Engines\/}}, ,  \vol{vol.~1},  \pg{pp. 58--79}.

\bibitem[Druault {\em et~al.\/}(2005)Druault, Delville \& Bonnet]{druault2005experimental}
{\sc \au{Druault, Philippe}, \au{Delville, Jo{\"e}l} \& \au{Bonnet, Jean-Paul}} \yr{2005}  \at{Experimental 3d analysis of the large scale behaviour of a plane turbulent mixing layer}.  \jt{Flow, Turbulence and Combustion}  \bvol{74},  \pg{207--233}.

\bibitem[D’Ovidio \& Coats(2013)]{d2013organized}
{\sc \au{D’Ovidio, A} \& \au{Coats, CM}} \yr{2013}  \at{Organized large structure in the post-transition mixing layer. part 1. experimental evidence}.  \jt{Journal of Fluid Mechanics}  \bvol{737},  \pg{466--498}.

\bibitem[Eaton(1995)]{eaton1995effects}
{\sc \au{Eaton, John~K}} \yr{1995}  \at{Effects of mean flow three dimensionality on turbulent boundary-layer structure}.  \jt{AIAA journal}  \bvol{33}~(11),  \pg{2020--2025}.

\bibitem[Fiedler {\em et~al.\/}(1998)Fiedler, Nayeri, Spieweg \& Paschereit]{fiedler1998three}
{\sc \au{Fiedler, HE}, \au{Nayeri, C}, \au{Spieweg, R} \& \au{Paschereit, CO}} \yr{1998}  \at{Three-dimensional mixing layers and their relatives}.  \jt{Experimental Thermal and Fluid Science}  \bvol{16}~(1-2),  \pg{3--21}.

\bibitem[Fiscaletti {\em et~al.\/}(2016)Fiscaletti, Attili, Bisetti \& Elsinga]{fiscaletti2016scale}
{\sc \au{Fiscaletti, Daniele}, \au{Attili, Antonio}, \au{Bisetti, Fabrizio} \& \au{Elsinga, Gerrit~E}} \yr{2016}  \at{Scale interactions in a mixing layer--the role of the large-scale gradients}.  \jt{Journal of Fluid Mechanics}  \bvol{791},  \pg{154--173}.

\bibitem[Fric(1996)]{fric1996skewed}
{\sc \au{Fric, TF}} \yr{1996}  \at{Skewed shear-layer mixing within a duct}.  \jt{AIAA journal}  \bvol{34}~(4),  \pg{847--849}.

\bibitem[Gatski \& Bonnet(2013)]{gatski2013compressibility}
{\sc \au{Gatski, Thomas~B} \& \au{Bonnet, Jean-Paul}} \yr{2013} {\em Compressibility, turbulence and high speed flow\/}.  \publ{Academic Press}.

\bibitem[G{\"o}rtler(1942)]{gortler1942berechnung}
{\sc \au{G{\"o}rtler, von~H}} \yr{1942}  \at{Berechnung von aufgaben der freien turbulenz auf grund eines neuen n{\"a}herungsansatzes.}  \jt{ZAMM-Journal of Applied Mathematics and Mechanics/Zeitschrift f{\"u}r Angewandte Mathematik und Mechanik}  \bvol{22}~(5),  \pg{244--254}.

\bibitem[Govardhan \& Dhanasekaran(2002)]{govardhan2002effect}
{\sc \au{Govardhan, M} \& \au{Dhanasekaran, TS}} \yr{2002}  \at{Effect of guide vanes on the performance of a self-rectifying air turbine with constant and variable chord rotors}.  \jt{Renewable Energy}  \bvol{26}~(2),  \pg{201--219}.

\bibitem[Gr{\"u}ndel \& Fiedler(1993)]{grundel1993mixing}
{\sc \au{Gr{\"u}ndel, H} \& \au{Fiedler, HE}} \yr{1993}  \at{The mixing layer between non-parallel streams}.  \jt{Applied Scientific Research}  \bvol{51},  \pg{167--171}.

\bibitem[Hackett \& Cox(1970)]{hackett1970three}
{\sc \au{Hackett, JE} \& \au{Cox, DK}} \yr{1970}  \at{The three-dimensional mixing layer between two grazing perpendicular streams}.  \jt{Journal of Fluid Mechanics}  \bvol{43}~(1),  \pg{77--96}.

\bibitem[Ho \& Huerre(1984)]{ho1984perturbed}
{\sc \au{Ho, C-M} \& \au{Huerre, Patrick}} \yr{1984}  \at{Perturbed free shear layers}.  \jt{Annual Review of Fluid Mechanics}  \bvol{16},  \pg{365--424}.

\bibitem[Jim{\'e}nez(2025)]{jimenez2025chaos}
{\sc \au{Jim{\'e}nez, Javier}} \yr{2025}  \at{Chaos, coherence, and turbulence}.  \jt{Physical Review Fluids}  \bvol{10}~(10),  \pg{100504}.

\bibitem[Jimenez {\em et~al.\/}(1985)Jimenez, Cogollos \& Bernal]{jimenez1985perspective}
{\sc \au{Jimenez, Javier}, \au{Cogollos, Marta} \& \au{Bernal, Luis~P}} \yr{1985}  \at{A perspective view of the plane mixing layer}.  \jt{Journal of Fluid Mechanics}  \bvol{152},  \pg{125--143}.

\bibitem[Johnston \& Flack(1996)]{johnston1996advances}
{\sc \au{Johnston, JP} \& \au{Flack, KA}} \yr{1996}  \at{Advances in three-dimensional turbulent boundary layers with emphasis on the wall-layer regions}.  \jt{Journal of Fluids Engineering} .

\bibitem[Kiesow \& Plesniak(2003)]{kiesow2003near}
{\sc \au{Kiesow, Robert~O} \& \au{Plesniak, Michael~W}} \yr{2003}  \at{Near-wall physics of a shear-driven three-dimensional turbulent boundary layer with varying crossflow}.  \jt{Journal of Fluid Mechanics}  \bvol{484},  \pg{1--39}.

\bibitem[Korpela(2019)]{korpela2019principles}
{\sc \au{Korpela, Seppo~A}} \yr{2019} {\em Principles of turbomachinery\/}.  \publ{John Wiley \& Sons}.

\bibitem[Kumar {\em et~al.\/}(2024)Kumar, Gupta, Bewley \& Larsson]{kumar2024three}
{\sc \au{Kumar, Vedant}, \au{Gupta, Dipendra}, \au{Bewley, Gregory~P} \& \au{Larsson, Johan}} \yr{2024} Three-dimensional effects in turbulent shear layers.  \bt{In {\em AIAA AVIATION FORUM AND ASCEND 2024\/}},  \pg{p. 4372}.

\bibitem[Laizet {\em et~al.\/}(2010)Laizet, Lardeau \& Lamballais]{laizet2010direct}
{\sc \au{Laizet, Sylvain}, \au{Lardeau, Sylvain} \& \au{Lamballais, Eric}} \yr{2010}  \at{Direct numerical simulation of a mixing layer downstream a thick splitter plate}.  \jt{Physics of Fluids}  \bvol{22}~(1).

\bibitem[Li {\em et~al.\/}(2010)Li, Chang \& Wang]{li2010experimental}
{\sc \au{Li, Chiuan-Ting}, \au{Chang, Keh-Chin} \& \au{Wang, Muh-Rong}} \yr{2010}  \at{Experimental study on evolution of joint velocity pdf in planar mixing layer}.  \jt{Experimental Thermal and Fluid Science}  \bvol{34}~(8),  \pg{1122--1132}.

\bibitem[Liepmann \& Laufer(1947)]{liepmann1947investigations}
{\sc \au{Liepmann, HW} \& \au{Laufer, J}} \yr{1947}  \bt{Investigations of free turbulent mixing}.  \org{{\em Tech. Rep.\/}}.

\bibitem[Littell \& Eaton(1994)]{littell1994turbulence}
{\sc \au{Littell, Howard~S} \& \au{Eaton, John~K}} \yr{1994}  \at{Turbulence characteristics of the boundary layer on a rotating disk}.  \jt{Journal of Fluid Mechanics}  \bvol{266},  \pg{175--207}.

\bibitem[Loucks \& Wallace(2012)]{loucks2012velocity}
{\sc \au{Loucks, Richard~B} \& \au{Wallace, James~M}} \yr{2012}  \at{Velocity and velocity gradient based properties of a turbulent plane mixing layer}.  \jt{Journal of Fluid Mechanics}  \bvol{699},  \pg{280--319}.

\bibitem[Lozano-Dur{\'a}n {\em et~al.\/}(2020)Lozano-Dur{\'a}n, Giometto, Park \& Moin]{lozano2020non}
{\sc \au{Lozano-Dur{\'a}n, Adri{\'a}n}, \au{Giometto, Marco~G}, \au{Park, George~Ilhwan} \& \au{Moin, Parviz}} \yr{2020}  \at{Non-equilibrium three-dimensional boundary layers at moderate reynolds numbers}.  \jt{Journal of Fluid Mechanics}  \bvol{883},  \pg{A20}.

\bibitem[Lu \& Lele(1999)]{lu1999asymptotic}
{\sc \au{Lu, G} \& \au{Lele, SK}} \yr{1999}  \at{Asymptotic growth of disturbances from spatially compact source in a skewed mixing layer}.  \jt{Physics of Fluids}  \bvol{11}~(5),  \pg{1153--1160}.

\bibitem[Lu \& Lele(1993)]{lu1993inviscid}
{\sc \au{Lu, Ganyu} \& \au{Lele, Sanjiva~K}} \yr{1993}  \at{Inviscid instability of a skewed compressible mixing layer}.  \jt{Journal of Fluid Mechanics}  \bvol{249},  \pg{441--463}.

\bibitem[Mahesh(2013)]{mahesh2013interaction}
{\sc \au{Mahesh, Krishnan}} \yr{2013}  \at{The interaction of jets with crossflow}.  \jt{Annual Review of Fluid Mechanics}  \bvol{45}~(1),  \pg{379--407}.

\bibitem[McMullan(2015)]{mcmullan2015spanwise}
{\sc \au{McMullan, William~A}} \yr{2015}  \at{Spanwise domain effects on the evolution of the plane turbulent mixing layer}.  \jt{International Journal of Computational Fluid Dynamics}  \bvol{29}~(6-8),  \pg{333--345}.

\bibitem[McMullan {\em et~al.\/}(2015)McMullan, Gao \& Coats]{mcmullan2015organised}
{\sc \au{McMullan, William~A}, \au{Gao, S} \& \au{Coats, Christopher~M}} \yr{2015}  \at{Organised large structure in the post-transition mixing layer. part 2. large-eddy simulation}.  \jt{Journal of Fluid Mechanics}  \bvol{762},  \pg{302--343}.

\bibitem[Mehta(1991)]{mehta1991effect}
{\sc \au{Mehta, RD}} \yr{1991}  \at{Effect of velocity ratio on plane mixing layer development: Influence of the splitter plate wake}.  \jt{Experiments in Fluids}  \bvol{10}~(4),  \pg{194--204}.

\bibitem[Mehta \& Bradshaw(1979)]{mehta1979design}
{\sc \au{Mehta, Ravi~Datt} \& \au{Bradshaw, Peter}} \yr{1979}  \at{Design rules for small low speed wind tunnels}.  \jt{The Aeronautical Journal}  \bvol{83}~(827),  \pg{443--453}.

\bibitem[Meldi {\em et~al.\/}(2020)Meldi, Mariotti, Salvetti \& Sagaut]{meldi2020numerical}
{\sc \au{Meldi, M}, \au{Mariotti, A}, \au{Salvetti, MV} \& \au{Sagaut, P}} \yr{2020}  \at{Numerical investigation of skewed spatially evolving mixing layers}.  \jt{Journal of Fluid Mechanics}  \bvol{897},  \pg{A35}.

\bibitem[Nygaard \& Glezer(1991)]{nygaard1991evolution}
{\sc \au{Nygaard, KJ} \& \au{Glezer, A}} \yr{1991}  \at{Evolution of stream wise vortices and generation of small-scale motion in a plane mixing layer}.  \jt{Journal of Fluid Mechanics}  \bvol{231},  \pg{257--301}.

\bibitem[Paschereit {\em et~al.\/}(1989)Paschereit, Sch{\"u}ttpelz \& Fiedler]{paschereit1989mixing}
{\sc \au{Paschereit, CO}, \au{Sch{\"u}ttpelz, M} \& \au{Fiedler, HE}} \yr{1989} The mixing layer between non-parallel walls.  \bt{In {\em Advances in Turbulence 2: Proceedings of the Second European Turbulence Conference, Berlin, August 30--September 2, 1988\/}},  \pg{pp. 467--471}. Springer.

\bibitem[Peacock {\em et~al.\/}(2024)Peacock, Pullan \& Folk]{peacock2024dissipation}
{\sc \au{Peacock, Robert}, \au{Pullan, Graham} \& \au{Folk, Masha}} \yr{2024} Dissipation in skewed boundary layers.  \bt{In {\em Turbo Expo: Power for Land, Sea, and Air\/}}, ,  \vol{vol. 88063},  \pg{p. V12BT30A027}. American Society of Mechanical Engineers.

\bibitem[Pirozzoli {\em et~al.\/}(2015)Pirozzoli, Bernardini, Mari{\'e} \& Grasso]{pirozzoli2015early}
{\sc \au{Pirozzoli, Sergio}, \au{Bernardini, Matteo}, \au{Mari{\'e}, Simon} \& \au{Grasso, Francesco}} \yr{2015}  \at{Early evolution of the compressible mixing layer issued from two turbulent streams}.  \jt{Journal of Fluid Mechanics}  \bvol{777},  \pg{196--218}.

\bibitem[Pope(2000)]{Pope_2000}
{\sc \au{Pope, Stephen~B.}} \yr{2000} {\em Turbulent Flows\/}.  \publ{Cambridge University Press}.

\bibitem[Prandtl(1942)]{prandtl1942bemerkungen}
{\sc \au{Prandtl, Ludwig}} \yr{1942}  \at{Bemerkungen zur theorie der freien turbulenz.}  \jt{ZAMM-Journal of Applied Mathematics and Mechanics/Zeitschrift f{\"u}r Angewandte Mathematik und Mechanik}  \bvol{22}~(5),  \pg{241--243}.

\bibitem[Riley \& Lowson(1998)]{riley1998development}
{\sc \au{Riley, AJ} \& \au{Lowson, MV}} \yr{1998}  \at{Development of a three-dimensional free shear layer}.  \jt{Journal of Fluid Mechanics}  \bvol{369},  \pg{49--89}.

\bibitem[Rogers \& Moser(1994)]{rogers1994direct}
{\sc \au{Rogers, Michael~M} \& \au{Moser, Robert~D}} \yr{1994}  \at{Direct simulation of a self-similar turbulent mixing layer}.  \jt{Physics of Fluids}  \bvol{6}~(2),  \pg{903--923}.

\bibitem[Roshko(2005)]{roshko2005plane}
{\sc \au{Roshko, A}} \yr{2005} The plane mixing layer flow visualization results and three dimensional effects.  \bt{In {\em The Role of Coherent Structures in Modelling Turbulence and Mixing: Proceedings of the International Conference Madrid, Spain, June 25--27, 1980\/}},  \pg{pp. 208--217}. Springer.

\bibitem[Sabin(1965)]{sabin1965analytical}
{\sc \au{Sabin, Cullen~Milo}} \yr{1965}  \at{An analytical and experimental study of the plane, incompressible, turbulent free-shear layer with arbitrary velocity ratio and pressure gradient}.  \jt{Journal of Basic Engineering} .

\bibitem[Sandham \& Sandberg(2009)]{Sandham_JT_2009}
{\sc \au{Sandham, N~D} \& \au{Sandberg, R~D}} \yr{2009}  \at{Direct numerical simulation of the early development of a turbulent mixing layer downstream of a splitter plate}.  \jt{Journal of Turbulence} ~(10),  \pg{N1}.

\bibitem[Schr{\"o}der {\em et~al.\/}(2000)Schr{\"o}der, Hofmann \& Hourmouziadis]{schroder2000trailing}
{\sc \au{Schr{\"o}der, N}, \au{Hofmann, G} \& \au{Hourmouziadis, J}} \yr{2000} Trailing edge 3d free shear layers.  \bt{In {\em Turbo Expo: Power for Land, Sea, and Air\/}}, ,  \vol{vol. 78545},  \pg{p. V001T03A012}. American Society of Mechanical Engineers.

\bibitem[Sharan {\em et~al.\/}(2019)Sharan, Matheou \& Dimotakis]{sharan2019turbulent}
{\sc \au{Sharan, Nek}, \au{Matheou, Georgios} \& \au{Dimotakis, Paul~E}} \yr{2019}  \at{Turbulent shear-layer mixing: initial conditions, and direct-numerical and large-eddy simulations}.  \jt{Journal of Fluid Mechanics}  \bvol{877},  \pg{35--81}.

\bibitem[Smits {\em et~al.\/}(2011)Smits, McKeon \& Marusic]{smits2011high}
{\sc \au{Smits, Alexander~J}, \au{McKeon, Beverley~J} \& \au{Marusic, Ivan}} \yr{2011}  \at{High--reynolds number wall turbulence}.  \jt{Annual Review of Fluid Mechanics}  \bvol{43}~(1),  \pg{353--375}.

\bibitem[Townsend(1976)]{townsend1976structure}
{\sc \au{Townsend, AAR}} \yr{1976} {\em The structure of turbulent shear flow\/}.  \publ{Cambridge university press}.

\bibitem[Yoder {\em et~al.\/}(2015)Yoder, DeBonis \& Georgiadis]{yoder2015modeling}
{\sc \au{Yoder, Dennis~A}, \au{DeBonis, James~R} \& \au{Georgiadis, Nicholas~J}} \yr{2015}  \at{Modeling of turbulent free shear flows}.  \jt{Computers \& fluids}  \bvol{117},  \pg{212--232}.

\bibitem[Yoon \& Warhaft(1990)]{yoon1990evolution}
{\sc \au{Yoon, K} \& \au{Warhaft, Z}} \yr{1990}  \at{The evolution of grid-generated turbulence under conditions of stable thermal stratification}.  \jt{Journal of Fluid Mechanics}  \bvol{215},  \pg{601--638}.

\end{thebibliography}

\end{document}